\pdfoutput=1
\documentclass{article}
\usepackage{arxiv}

\usepackage{upgreek}
\usepackage{textcomp}
\usepackage{bm} 

\usepackage{amssymb, amsmath} 
\usepackage{float} 
\usepackage[utf8]{inputenc} % allow utf-8 input
\usepackage[T1]{fontenc}    % use 8-bit T1 fonts
\usepackage{url}            % simple URL typesetting
\usepackage{booktabs}       % professional-quality tables
\usepackage{amsfonts}       % blackboard math symbols
\usepackage{nicefrac}       % compact symbols for 1/2, etc.
\usepackage{microtype}      % microtypography
\usepackage{graphicx}
\usepackage{gensymb}

\title{Strain Induced Anisotropic Magnetic Behaviour and Exchange Coupling Effect in Fe-SmCo$_{5}$ Permanent Magnets Generated by High Pressure Torsion}

\author{
  Lukas Weissitsch \\
  Erich Schmid Institute of Materials Science, Austrian Academy of Sciences\\
8700 Leoben, Austria \\
  \texttt{lukas.weissitsch@oeaw.ac.at}
\And
     Martin St\"uckler \\
  Erich Schmid Institute of Materials Science, Austrian Academy of Sciences\\
8700 Leoben, Austria \\
\And
   Stefan Wurster \\
     Erich Schmid Institute of Materials Science, Austrian Academy of Sciences\\
8700 Leoben, Austria\\
  \And
   Peter Knoll \\
     Institute of Physics\\
  University of Graz, 8010 Graz, Austria\\ 
    \And
   Heinz Krenn \\
     Institute of Physics\\
  University of Graz, 8010 Graz, Austria\\ 
\And
  Reinhard Pippan\\
    Erich Schmid Institute of Materials Science, Austrian Academy of Sciences\\
8700 Leoben, Austria 
\And
  Andrea Bachmaier\\
  Erich Schmid Institute of Materials Science, Austrian Academy of Sciences\\
8700 Leoben, Austria 
}

\begin{document}

\maketitle

\begin{abstract}
High-pressure torsion (HPT), a technique of severe plastic deformation (SPD), is shown as a promising processing method for exchange-spring magnetic materials in bulk form. Powder mixtures of Fe and SmCo$_{5}$ are consolidated and deformed by HPT exhibiting sample dimensions of several millimetres, being essential for bulky magnetic applications. The structural evolution during HPT deformation of Fe-SmCo$_{5}$ compounds at room- and elevated- temperatures of chemical compositions consisting of 87, 47, 24 and 10~wt.\% Fe is studied and microstructurally analysed. Electron microscopy and synchrotron X-ray diffraction reveal a dual-phase nanostructured composite for the as-deformed samples with grain refinement after HPT deformation.
SQUID magnetometry measurements show hysteresis curves of an exchange coupled nanocomposite at room temperature, while for low temperatures a decoupling of Fe and SmCo$_{5}$ is observed. Furthermore, exchange interactions between the hard- and soft-magnetic phase can explain a shift of the hysteresis curve. Strong emphasis is devoted to the correlation between the magnetic properties and the evolving nano-structure during HPT deformation, which is conducted for a 1:1 composition ratio of Fe to SmCo$_{5}$. SQUID magnetometry measurements show an increasing saturation magnetisation for increasing strain $\gamma$ and a maximum of the coercive field strength at a shear strain of $\gamma$~=~75.
\end{abstract}

% Keywords

%%%%%%%%%%%%%%%%%%%%%%%%%%%%%%%%%%%%%%%%%%
%%%%%%%%%%%%%%%%%%%%%%%%%%%%%%%%%%%%%%%%%%

\section{Introduction}

Permanent magnets are designed for a high saturation magnetisation, a high Curie temperature and a high anisotropy resulting in a strong coercivity and yielding a high energy product. The maximal achievable energy product of a magnet is a widely used and crucial parameter to compare different magnetic materials or processing routes.
Compared to Nd-Fe-B based magnets, the intermetallic Sm-Co compounds offer a relatively low saturation magnetisation. Sm-Co based magnets instead provide the highest Curie temperatures (up to 1190~K), which is essential for applications at elevated temperatures \cite{ray1975}. State-of-the-art magnetic materials, which provide a maximum magnetic field, require supplementary cooling, reducing the overall efficiency \cite{rong2011}.
Moreover, the demand for permanent magnetic materials is strongly increasing, no matter if implemented in electric motors for vehicles or in windmills producing renewable energy. The achievable magnetic properties of common magnetic materials are already close to theoretical predictions \cite{gutfleisch2000}. Furthermore, mining and production processes of rare-earth elements are often correlated to political instabilities in the countries of origin and to monopolistic situations, which influences the price of elements on the world market \cite{gutfleisch2011}. 

The promising approach of exchange coupled spring magnets as permanent magnetic materials is an interesting field for research. The concept, first introduced nearly 30 years ago by Kneller and Hawig~\cite{kneller1991}, combines the high remanence and saturation magnetisation of a soft-magnet and the huge coercive field strength of a hard magnet.
Therefore the energy product of a magnet is increased, when the required magnetic anisotropy, provided by the hard magnetic phase, stabilises the coupled soft magnetic phase against demagnetisation. Soon after, the influence of the morphology, such as feature size and geometry between hard and soft magnetic phases, on the resulting hysteresis loop was studied. According to theoretical considerations, energy products higher than 120 MGOe, which is considerably higher than present high performance hard magnets (about 50--60~MGOe), should be achieved \cite{skomski1993,sahota2012, gutfleisch2011}. 
The concept further predicts, that the dimension of the soft-magnetic phase has to be smaller than the exchange length of the hard-magnetic phase, which necessitates nanostructuring. 
First experiments were conducted on well tuneable thin films and another techniques have been used, such as core-shell structured nanoparticles, rapidly quenched ribbons or mechanically milled powders \cite{fullerton1999}. Despite important contributions on fundamental knowledge, up-scaling and processing of bulk samples with an internal nanostructure is nowadays a challenging task. 

Nanostructured materials are commonly produced by bottom-up methodologies, which compulsorily requires a consolidation step of powder precursors as bulk solid magnets are desired for practical applications. Depending on the technological process, the magnets suffer from a weak particle bonding, which favours mechanical instabilities as well as corrosion vulnerability. A residual porosity diminishes the bulk magnetic properties and thermal treatment during a sintering process lead to grain growth, which is another limiting factor. 
Severe plastic deformation (SPD) by high-pressure torsion (HPT) is a top-down approach to obtain nanostructured materials, i.e., grain refinement is induced in a bulk material upon processing. Thus, common processing limits can be avoided and bulk samples in millimetre sized dimensions are obtained, which have been recently increased to the 100~mm regime \cite{valiev2000, hohenwarter2009, edalati2016, hohenwarter2019}.
In this study, we demonstrate the feasibility of HPT-deformation to process exchange coupled spring magnets. In the materials investigated, Fe is used as soft magnetic phase and SmCo$_{5}$ as hard magnetic phase. The successful fabrication and microstructural characterization of a dual-phase nanostructured composite is described in detail and resulting magnetic properties are measured. Additionally, the dependency of magnetic properties on the applied strain during HPT-deformation is investigated.

%%%%%%%%%%%%%%%%%%%%%%%%%%%%%%%%%%%%%%%%%%
\section{Materials and Methods}

The HPT-deformation procedure is best described as a severe shearing process by torsion. In an idealised process the applied shear $\gamma $ is defined by 

\begin{equation} \label{eq:gamma}
\gamma = \dfrac{2 \pi r}{t} n 
\end{equation}
as a function of the radius \textit{r} within a disc subjected to \textit{n} revolutions. The sample thickness \textit{t} is considered to be constant. The used HPT setup is described in detail in ref. \cite{hohenwarter2009}.

As starting materials conventional powders (Fe: MaTeck 99.9\% $-$ 100 + 200 mesh; SmCo$_{5}$: Alfa Aesar, intermetallic fine powder), which are stored and handled in an Ar filled glove box, are used to obtain the desired chemical composition. The investigated compounds consist of 10, 24, 47 and 87~wt.\% soft magnetic Fe and 90, 76, 53 and 13~wt.\% hard magnetic SmCo$_{5}$, respectively. Weighing the starting powders is very precise about an error below $\pm$0.15~wt.\%. Additional EDX measurements along the cross-section on final samples, reveal an over-all-error of less than $\pm$1.5~wt.\%. 
This does not hold for smaller specimen extracted at high applied strains exhibiting a separation into two phases: Here the local chemical composition varies to a far greater extent and does not exactly average to an exact global sample composition.
The mixed powders are hydrostatically compacted under Ar-atmosphere using the HPT device at a nominal pressure of 5~GPa and a quarter revolution. Shortly thereafter, the samples are severely deformed at 7.5~GPa with a rotational frequency of 0.6~min$^{-1}$ to avoid a significant increase of the processing temperature, which is further suppressed by high pressure air cooling. These samples are denoted as room-temperature (RT) deformed samples in the following.
No prediction of the deformation behaviour of composites can be made on simple basis of the corresponding material properties \cite{kormout2017, pippan2010}.
During HPT-deformation at RT crack formation, a higher sample hardness and a limited amount of revolutions is not avoidable for certain compositions. To apply a higher amount of strain while avoiding crack formation, the influence of deformation temperature on the deformation behaviour and phase evolution, is investigated. Hence, HPT-deformation at elevated temperatures (250  $^{\circ}$C
 and 400  $^{\circ}$C
) is carried out for selected samples.

\begin{figure}[H]
\centering
\includegraphics[width=0.6\textwidth]{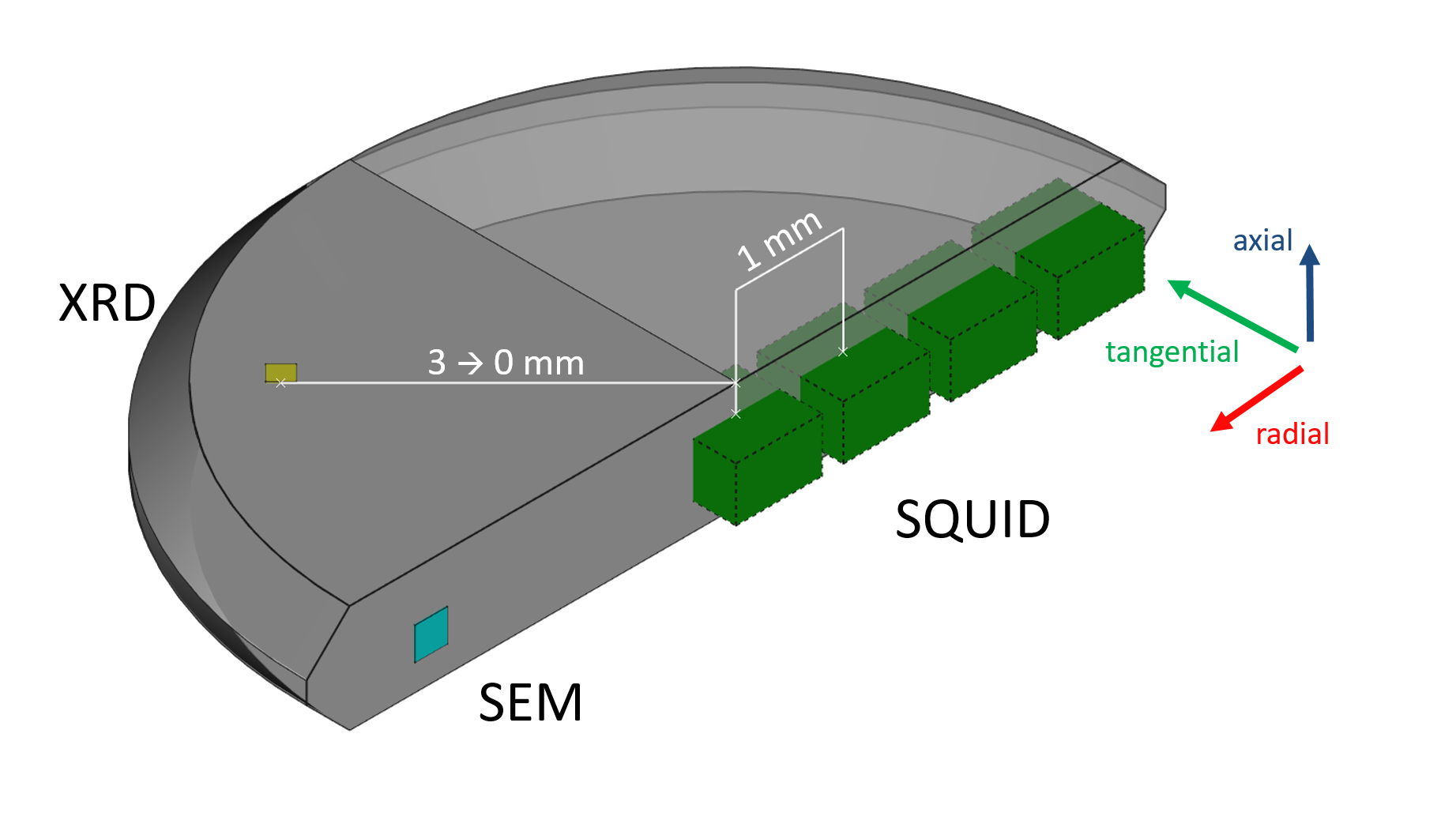}
\caption{Schematic illustration of a halved HPT-deformed disc. The positions of SQUID-samples at varying radii are outlined (green boxes) with respect to the used cutting directions. Exemplarily axial XRD (yellow) and tangential SEM (turquoise) measurements are indicated.}
  \label{fig:schema}
\end{figure}

After deformation the samples are cut into pieces for specific investigations. A schematic drawing of an as-deformed sample is depicted in Figure \ref{fig:schema}. 
The excised SQUID samples are displayed as well positions and directions, where measurements are carried out. 
Scanning electron microscopy (SEM; LEO 1525,  Carl Zeiss Microscopy GmbH, Oberkochen, Germany) images are made in tangential direction as well as the backscattered electron detection mode (BSE). Chemical evaluations are obtained with energy dispersive X-ray spectroscopy (EDX; XFlash 6--60, Bruker, Berlin, Germany) while using the software package Esprit 2.2 from Bruker. For line-scans, an acceleration voltage of 3--5~keV is used, resulting in a calculated maximal excitation volume radius of 70~nm and the obtained data are averaged over 7 pixels for smoothing. Vickers hardness measurements (Micromet 5104, Buehler, Lake Bluff, IL, USA) are carried out in tangential direction on the polished sample along the diameter in steps of $ \Delta$r~=~0.25~mm. A mean value is calculated if the measured microhardness saturates in a plateau (usually valid for: 2.0~<~r~<~3.5~mm). 
Synchrotron X-ray diffraction (XRD) measurements are carried out in axial orientation and transmission mode (Petra III: P07 synchrotron facility at Deutsches Elektronen-Synchrotron DESY, Hamburg, Germany; Beam Energy: 100~keV; Beam Size: 0.2 $\times$ 0.2~mm$^2$). The signals are calibrated with a CeO$_{2}$ reference and integrated with pyFAI. The processing software Fityk is subsequently used for peak analysis. As theoretical reference peaks the Crystallography Open Database (COD) is used (SmCo$_{5}$: COD 1524132, Fe: COD 9006587).

A SQUID-Magnetometer (Quantum Design MPMS-XL-7,Quantum Design,  Inc.,  San Diego,  CA, USA) operated with the manufacturer’s software MPMSMultiVu Application (version 1.54) is used for magnetic measurements. The hysteresis is measured at different temperatures in magnetic fields up to 70~kOe. Magnetic measurements are exclusively performed at RT deformed samples. The magnetic field is applied in axial HPT-disc orientation, if not other stated. For strain dependent measurements several SQUID samples are cut out of one HPT-disc (see Figure \ref{fig:schema}), whereas different radii can be linked to a specific amount of strain according to Equation (\ref{eq:gamma}). 
Due to finite sample sizes, rather a range of radii is observed instead of one specific radius. The investigated samples still represent bulk dimensional material, meaning the applied strain is averaged within radii of $\pm$~0.5~mm.

In Table \ref{tab:samples}, the processing parameters of all used samples in this study are summarised. Their chemical composition, deformation temperature \textit{def.-temp}, number of HPT-revolutions \textit{n} and resulting strain \textit{$\gamma $} at r~=~3~mm, according to Equation (\ref{eq:gamma}), is given. 

\begin{table}[H]
	\caption{Summarised processing parameters of used samples in this study. \textit{n} depicts the number of torsions and $\gamma $(r3) the shear strain at radius = 3~mm.}\label{tab:samples}
    \centering
    \begin{tabular}{cccccc}
    \toprule
        \textbf{Nr }	& \textbf{Fe/wt.}\bm{\% }& \textbf{SmCo}$_{\boldsymbol{5}}$\textbf{/wt.}\bm{\% } & \textbf{Def.-Temp} 	& $\boldsymbol{n}$ 	& $\boldsymbol{\gamma} $\textbf{(r3)} \\ \midrule
        1 	& 10 		& 90 				& RT 			& 25 	& 396 \\ 
        2 	& 24 		& 76 				& RT 			& 25 	& 445 \\ 
        3 	& 47 		& 53 				& RT 			& 7 	& 111 \\
        4 	& 87 		& 13 				& RT 			& 14 	& 206 \\
        5 	& 47 		& 53 				& 250 $^{\circ}$C
 		& 8 	& 118 \\ 
        6 	& 47 		& 53 				& 400  $^{\circ}$C
 		& 18 	& 265 \\ 
 		\bottomrule
    \end{tabular}
\end{table}

%%%%%%%%%%%%%%%%%%%%%%%%%%%%%%%%%%%%%%%%%%
\section{Results}

%%%%%%%%%%%%%%%%%%%%%%%%%%%%%%%%%%%%%%%%%%
\subsection{Influence of Processing Parameters on Structural Evolution}
\label{sec:microstructure}

\begin{figure}[H]
\centering
\includegraphics[width=13 cm]{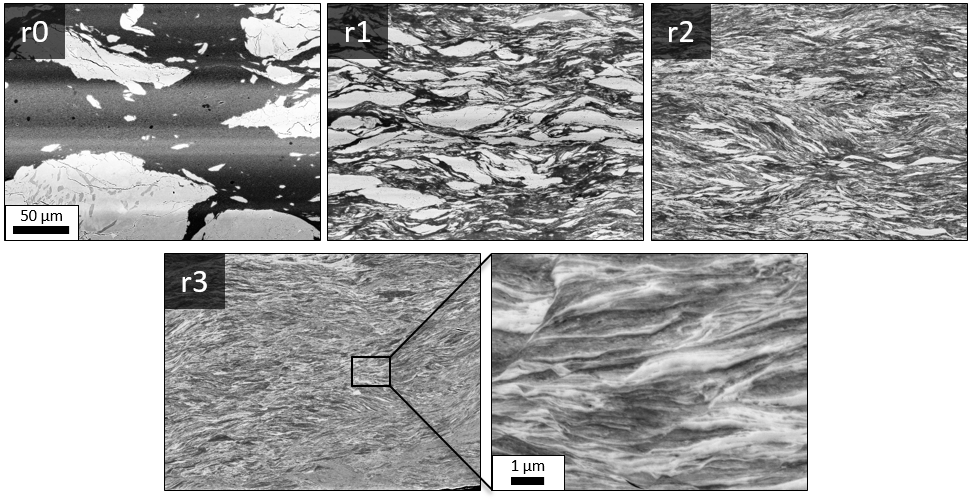}
\caption{BSE images at the same magnification of a HPT-deformed sample at RT consisting of 47~wt.\%~Fe--53~wt.\% SmCo$_{5}$ at different radii. The higher magnified image is taken at r =  3~mm.}
  \label{fig:microstructure_radial}
\end{figure}

In Figure \ref{fig:microstructure_radial}, the microstructure of the 47 wt.\% Fe--53 wt.\% SmCo$_{5}$ sample HPT-deformed at 7 revolutions at RT is shown for different positions along the HPT-disc. For all applied strains, a heterostructure is observed. In the center of the HPT-disc, at radius about 0, the lowest amount of shear strain ($\gamma$ $\sim$ 0 ) is applied and the material is mainly deformed by compression. BSE images show an inhomogeneous structure with separated phases of Fe (dark contrast) and SmCo$_{5}$ (bright contrast). According to Equation (\ref{eq:gamma}), the applied strain increases with increasing radius: $\gamma$ $\sim$ 37, $\sim$75 and $\sim$111 for r1, r2, r3, respectively. Plastic deformation is mainly observed in the Fe phase, limited plastic deformation, a splitting and localized shearing of the SmCo$_{5}$ phases is found. Nonetheless, a refinement of the structure is noticed with increasing strain. This behaviour during HPT-deformation is also observed for all other investigated chemical compositions (see Appendix \ref{sec:appendix}). The magnified image at r3 shows that the microstructural features become very small and finely dispersed, but strong contrast changes in the BSE image reveal still separated Fe and SmCo$_{5}$ phases. The phases are elongated in radial direction, but offer a thickness well below 1 $\upmu$m in axial direction. At the applied pressure it was not possible to fabricate samples with a higher strain due to slipping between sample and anvils. The sample hardness saturates at 648 $\pm$ 15~HV.

By varying the Fe content in the initial powder mixtures, a larger amount of strain can be applied. Figure \ref{fig:microstructure,chemical} shows the microstructure at r3 of three different compositions at two different magnifications. 
The sample consisting of 10 wt.\% Fe--90 wt.\% SmCo$_{5}$ is depicted in Figure \ref{fig:microstructure,chemical}a) after 25 HPT revolutions. Strong contrast differences reveal the bright SmCo$_{5}$ matrix with small Fe rich areas at a strain of $\gamma$ $\sim$ 396. The radial elongation of the Fe phase is in the range of a few $\upmu$m, while the axial dimension is significantly below 1 $\upmu$m.
Figure \ref{fig:microstructure,chemical}b) shows the microstructure of the sample consisting of 24 wt.\% Fe--76 wt.\% SmCo$_{5}$, HPT-deformed for 25 revolutions corresponding to a strain of $\gamma$ $\sim$ 445. At lower magnifications the absence of contrast differences would indicate a homogeneous microstructure, which is not confirmed when looking at higher magnifications. A lamellar morphology, well below 1 $\upmu$m is still visible. A similarity to the microstructure of the sample consisting of 47 wt.\% Fe--53 wt.\% SmCo$_{5}$ (Figure \ref{fig:microstructure_radial}) is noticed.
A sample with further increased Fe content (87 wt.\% Fe--13 wt.\% SmCo$_{5}$) is shown in Figure \ref{fig:microstructure,chemical}c). 14 HPT rotations are applied which corresponds to a strain of $\gamma$ $\sim$ 206. The microstructure shows a Fe matrix with finely elongated SmCo$_{5}$ phases. A further deformation is limited by an immense hardness increase leading to crack formation at the used HPT conditions.
The microhardness of the samples in Figure \ref{fig:microstructure,chemical}a--c) reaches 616~$\pm$~15, 657 $\pm$ 5 and 833 $\pm$ 11~HV, respectively.

\begin{figure}[H]
	\centering
	\includegraphics[width=13 cm]{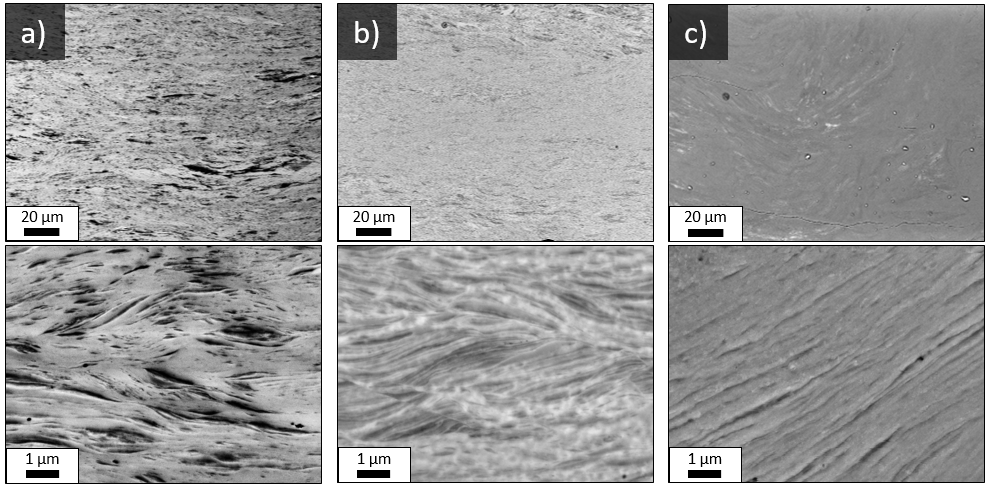}
	\caption{BSE images with two magnifications of HPT-deformed (RT) Fe-SmCo$_{5}$ samples at r = 3 mm consisting of (\textbf{a}) 10, (\textbf{b}) 24 and (\textbf{c}) 87 wt.\% Fe.}
	  \label{fig:microstructure,chemical}
\end{figure}

To investigate the lamellar microstructure in more detail, an EDX line-scan is carried out for one composition (47 wt.\% Fe--53 wt.\% SmCo$_{5}$). The scan direction is performed in axial HPT-disc direction, crossing the phases perpendicular as can be seen in the image of the microstructure in Figure~\ref{fig:microstructure,linescan}a) (top row). Strong changes of the Fe content can be explained by sharp interfaces between the Fe and SmCo$_{5}$ phase (Figure \ref{fig:microstructure,linescan}). 
Because the excitation volume of the electron beam exceeds the dimension size of the phases, a smoothening occurs. Thus, the interaction volume of the electron beam always interacts with both phases suppressing 100\%-Fe peaks. 
The ratio between Co and Sm content is calculated and analysed by a log-normal distribution which exhibits a peak at 4.84 $\pm$ 0.36. This closely corresponds to the stoichiometric distribution of the starting material of SmCo$_{5}$ and indicates that the phases remain unchanged during deformation and refinement. 
Similar line-scans are performed on the sample deformed at 400 $^{\circ}$C
. Here an increased maximum of the distribution is found pointing towards a higher Co to Sm ratio (e.g., Sm$_{2}$Co$_{17}$).

\begin{figure}[H]
	\centering
	\includegraphics[width=0.49\textwidth]{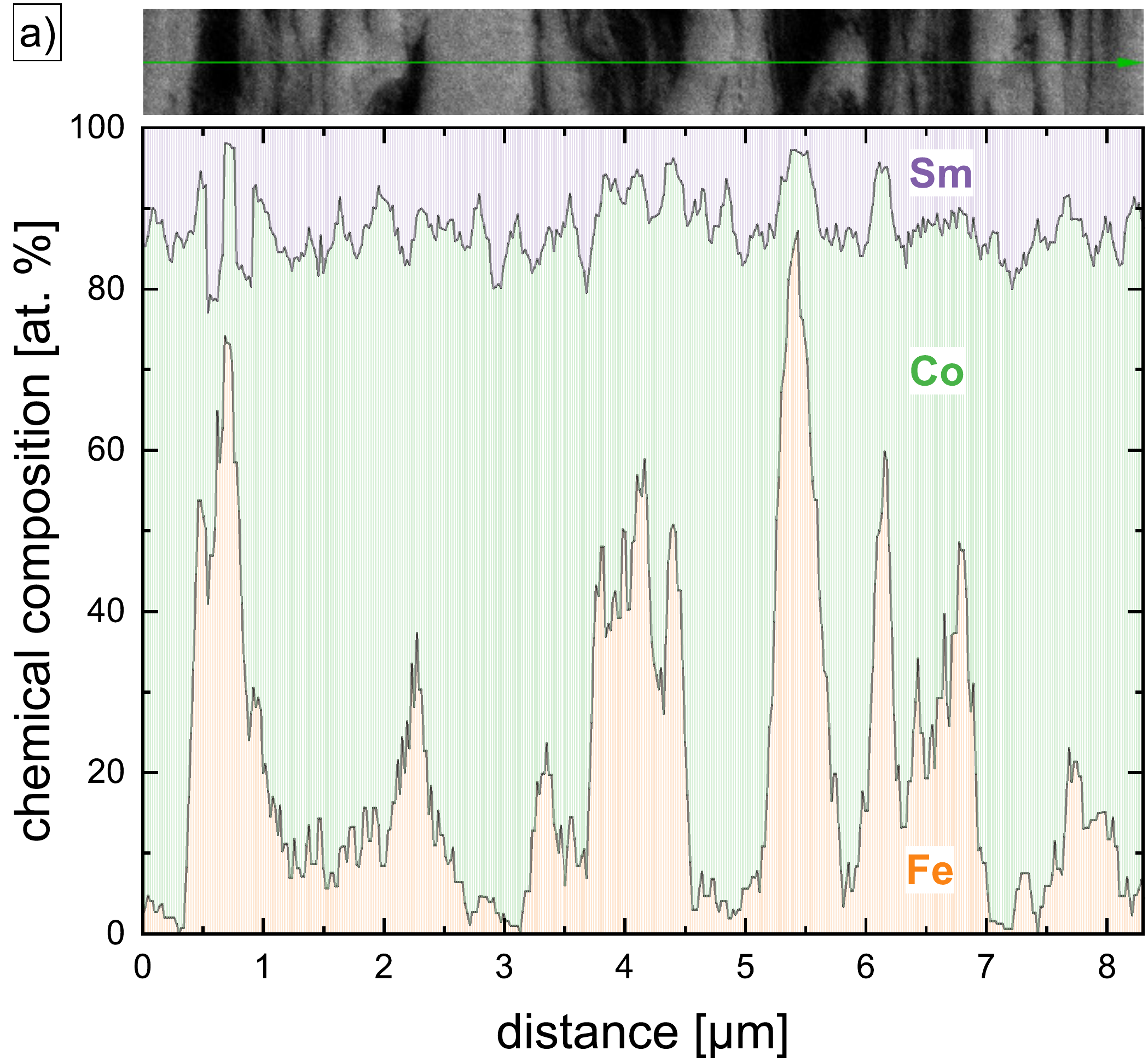}
	\includegraphics[width=0.49\textwidth]{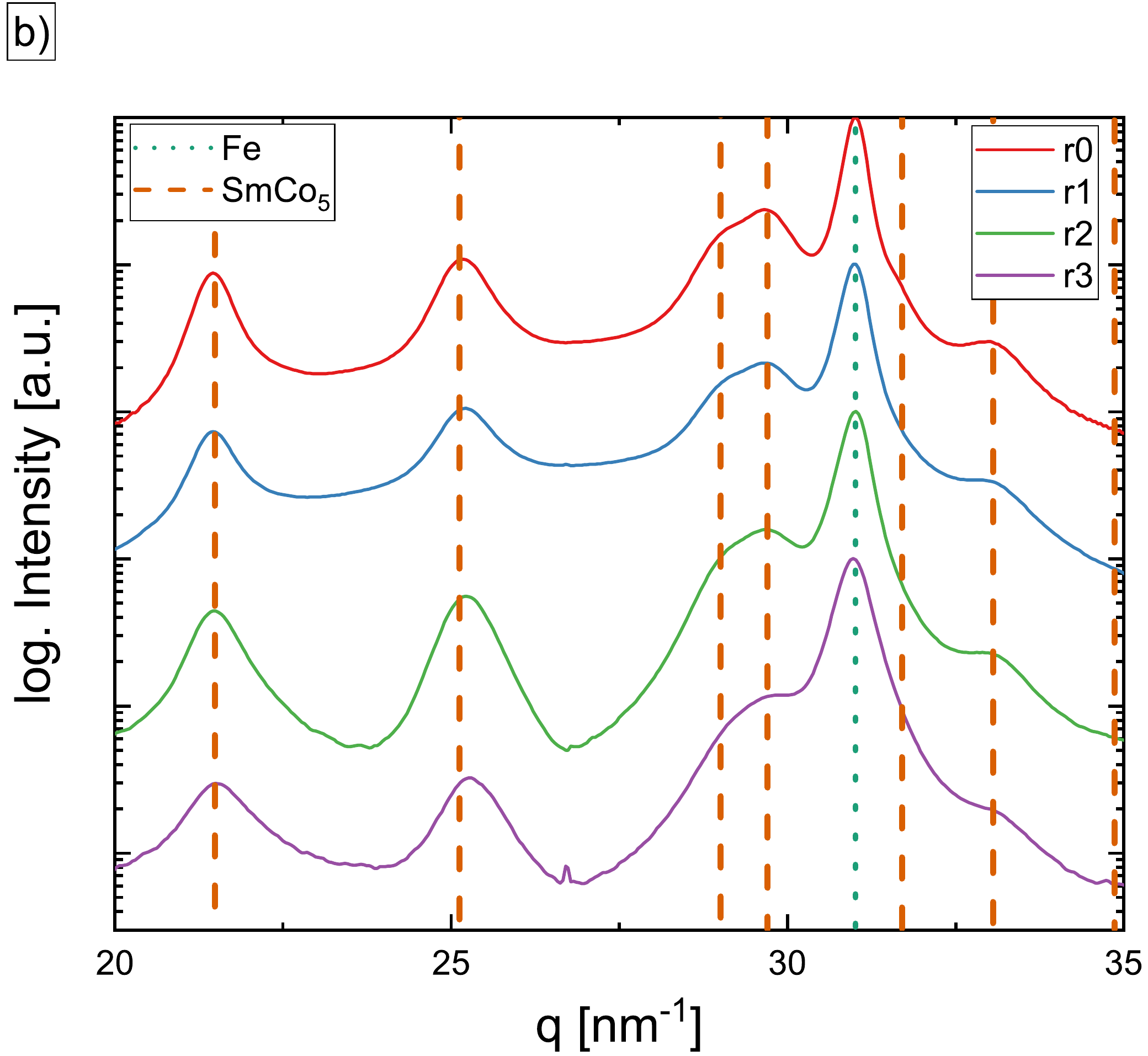}
	\caption{(\textbf{a}) EDX line-Scan at r3 in axial direction for the 47 wt.\% Fe--53 wt.\% SmCo$_{5}$ sample shown in Figure \ref{fig:microstructure_radial}. The alternating Fe content represents the two-phase microstructure. (\textbf{b}) Synchrotron XRD measurements of the RT-deformed 47 wt.\% Fe--53 wt.\% SmCo$_{5}$ sample for different radii.}
	  \label{fig:microstructure,linescan}
\end{figure}

Another proof for the stability of the phases during the deformation process are synchrotron XRD measurements. In Figure \ref{fig:microstructure,linescan}b) measurements of the 47 wt.\% Fe--53 wt.\% SmCo$_{5}$ sample along the radius corresponding to the positions of the BSE images in Figure \ref{fig:microstructure_radial} are presented. The theoretical reference peaks are in close accordance with the measured ones. With increasing radius, a peak broadening due to a grain size decrease is visible. The formation of additional phases cannot be observed.

Elevated deformation temperatures often helps to overcome processing limits, but usually lead to a final microstructure with larger grain sizes of the as-deformed state in the respective phases \cite{renk2019}.
Figure \ref{fig:DESY,temperature}a) shows synchrotron XRD-pattern for three Fe-SmCo$_{5}$ samples consisting of 47 wt.\% Fe--53 wt.\% SmCo$_{5}$, which were HPT-deformed at RT, 250 $^{\circ}$C
 and 400 $^{\circ}$C
. With increasing temperature, the peaks are getting sharper indicating coarser grains, but the Fe and SmCo$_{5}$ phases remain stable up to a deformation temperature of 250 $^{\circ}$C
. A strong difference is found at a deformation temperature of 400 $^{\circ}$C
. Several peaks are formed which cannot be related to the SmCo$_{5}$ phase. The formation of Sm$_{2}$Co$_{7}$ and Sm$_{2}$Co$_{17}$ phases is observed and examples of corresponding peaks are highlighted by black (Sm$_{2}$Co$_{7}$) and blue (Sm$_{2}$Co$_{17}$) arrows. 
To investigate the influence of the amount of strain for phase formations under elevated deformation temperatures, the XRD-patterns of one 400 $^{\circ}$C-deformed sample are shown in Figure \ref{fig:DESY,temperature}b) as a function of radius. At r0 the peaks correspond to the theoretical references of Fe and SmCo$_{5}$. The formation of additional phases is observed at higher radii, thus, the results show that the decomposition of SmCo$_{5}$ into Sm$_{2}$Co$_{7}$ and Sm$_{2}$Co$_{17}$ is not taking place at higher deformation temperature alone. The structural transition starts at a strain of $\gamma \sim$ 90 (r1 - blue arrow in Figure \ref{fig:DESY,temperature} b) and is more pronounced at even higher amounts of strain. 

\begin{figure}[H]
	\centering
	\includegraphics[width=0.49\textwidth]{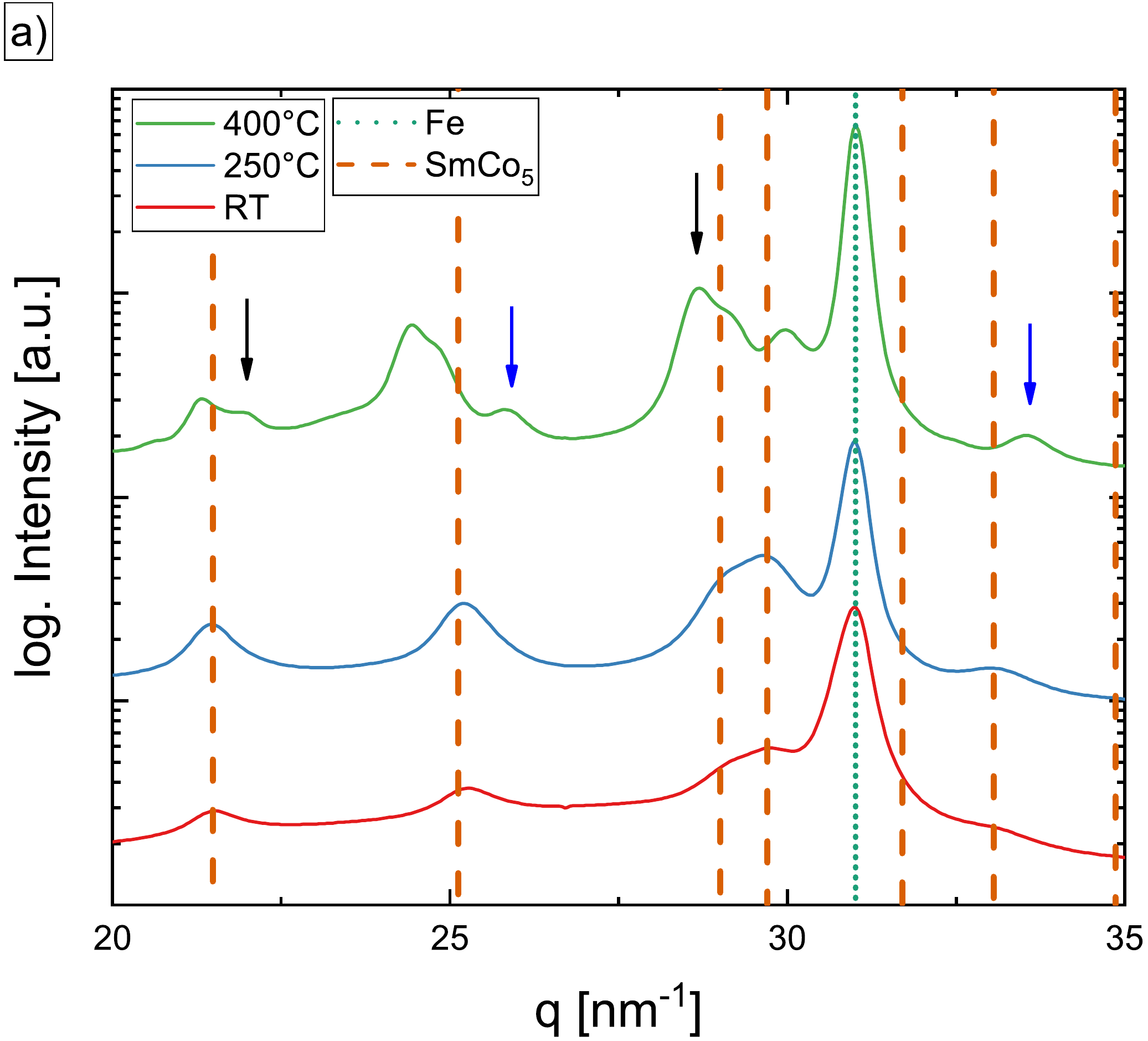}
	\includegraphics[width=0.49\textwidth]{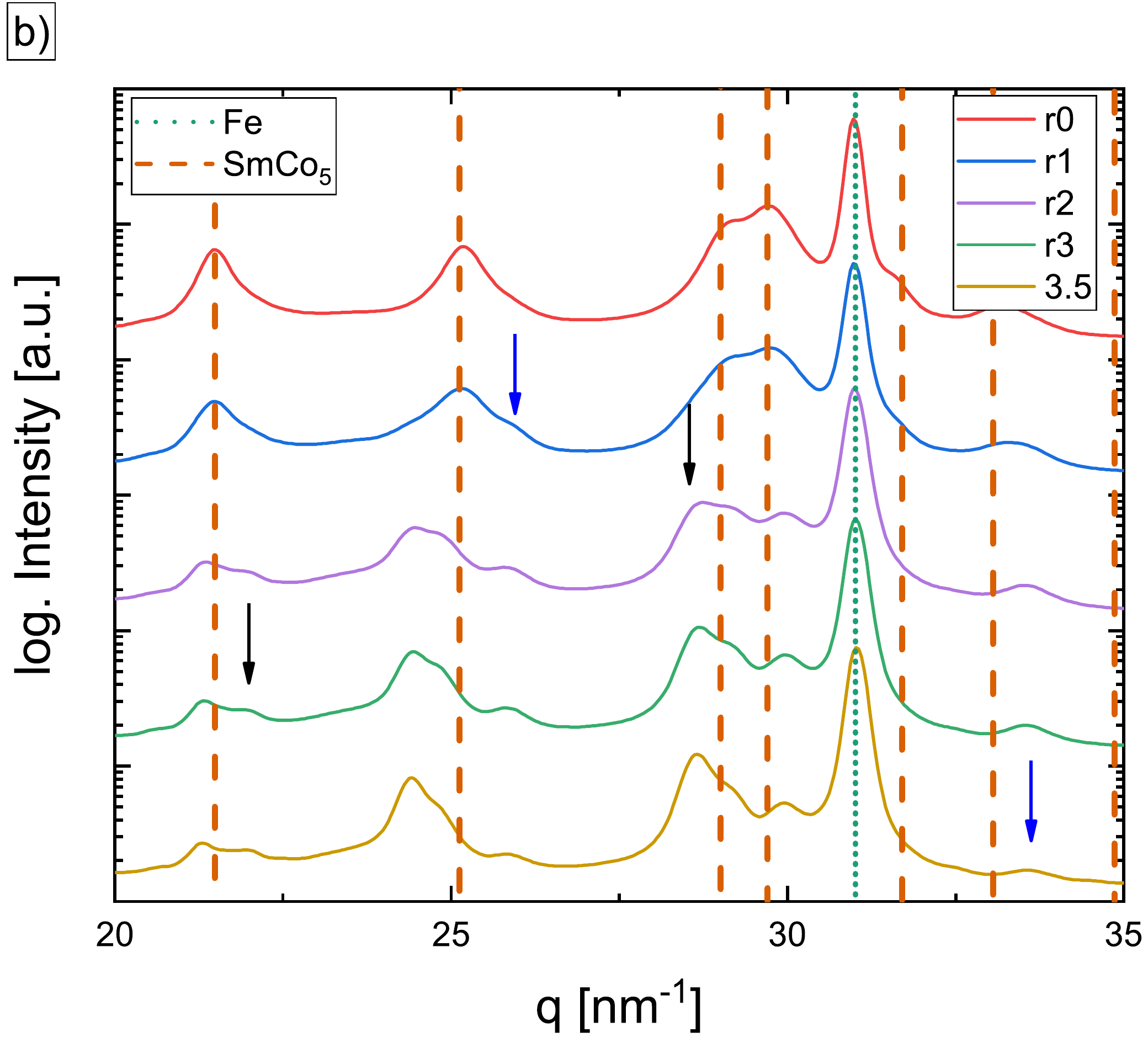}
	\caption{Synchrotron XRD measurements of several Fe-47 wt.\% samples deformed at different temperatures at r3 (\textbf{a}) and for the sample deformed at 400 $^{\circ}$C
 as a function of the radius (\textbf{b}). Examples of peak positions of the Sm$_{2}$Co$_{7}$ (black) and Sm$_{2}$Co$_{17}$ (blue) are indicated by arrows.}
  \label{fig:DESY,temperature}
\end{figure}

%%%%%%%%%%%%%%%%%%%%%%%%%%%%%%%%%%%%%%%%%%
\subsection{Magnetic Properties}

The magnetic properties of RT-deformed Fe-SmCo$_{5}$ compounds consisting of 87, 47, 24 and 10~wt.\% Fe are measured using a SQUID with the field applied in axial HPT-disc direction (see Figure \ref{fig:schema}). HPT-deformation induces a structure with lamellar morphology, where a polycrystalline microstructure is still present in both phases. Therefore, the applied magnetic field is perpendicular to this visible lamellar phase morphology. For every hysteresis measurement, a magnetic field up to $\pm$70~kOe is applied. In  Figure~\ref{fig:SQUID,chemical}, the measured magnetic moment is normalized to the mass of the sample and plotted versus the applied magnetic field. 
Hysteresis measurements at 8 K reveal decoupled hysteresis loops: The recoil curve (when the field is increased after demagnetisation) shows the behaviour of a superposition of a soft and a hard magnetic phase, i.e., the behaviour of an exchange-spring magnet without coupling. This is because of the formation of a `knee' due to two separated independent phases \cite{kneller1991}. Therefore, the soft magnetic $\alpha$-Fe phase is recognised by a high saturation magnetisation and a demagnetisation at low fields. The hard magnetic SmCo$_{5}$ phase is demagnetised at higher applied fields forming the `knee'. This effect can be seen in the hysteresis curve of all chemical compositions, but is stronger pronounced at higher SmCo$_{5}$ contents. 
At higher measurement temperatures, a coupling between soft and hard magnetic phases is clearly present, showing less constriction of the (more smooth) hysteresis curves.
Hysteresis measurements at 300~K (red curves) cannot be explained by a simple superposition of a separate Fe and SmCo$_{5}$ hysteresis. 
Here an exchange coupling is turned on, displaying an overall concave hysteresis profile with a concomitant decrease of coercivity.
The mass saturation magnetisation is decreasing with decreasing Fe content. In contrast, the value of the 8~K hysteresis lies slightly higher for all measurements. Although the maximal applied magnetic field of the hysteresis is 70~kOe, a complete saturation for the samples consisting of 47, 24 and 10~wt.\% Fe cannot be ensured, see Figure \ref{fig:SQUID,chemical}b--d).

\begin{figure}[H]
	\centering
	\includegraphics[width=0.4\textwidth]{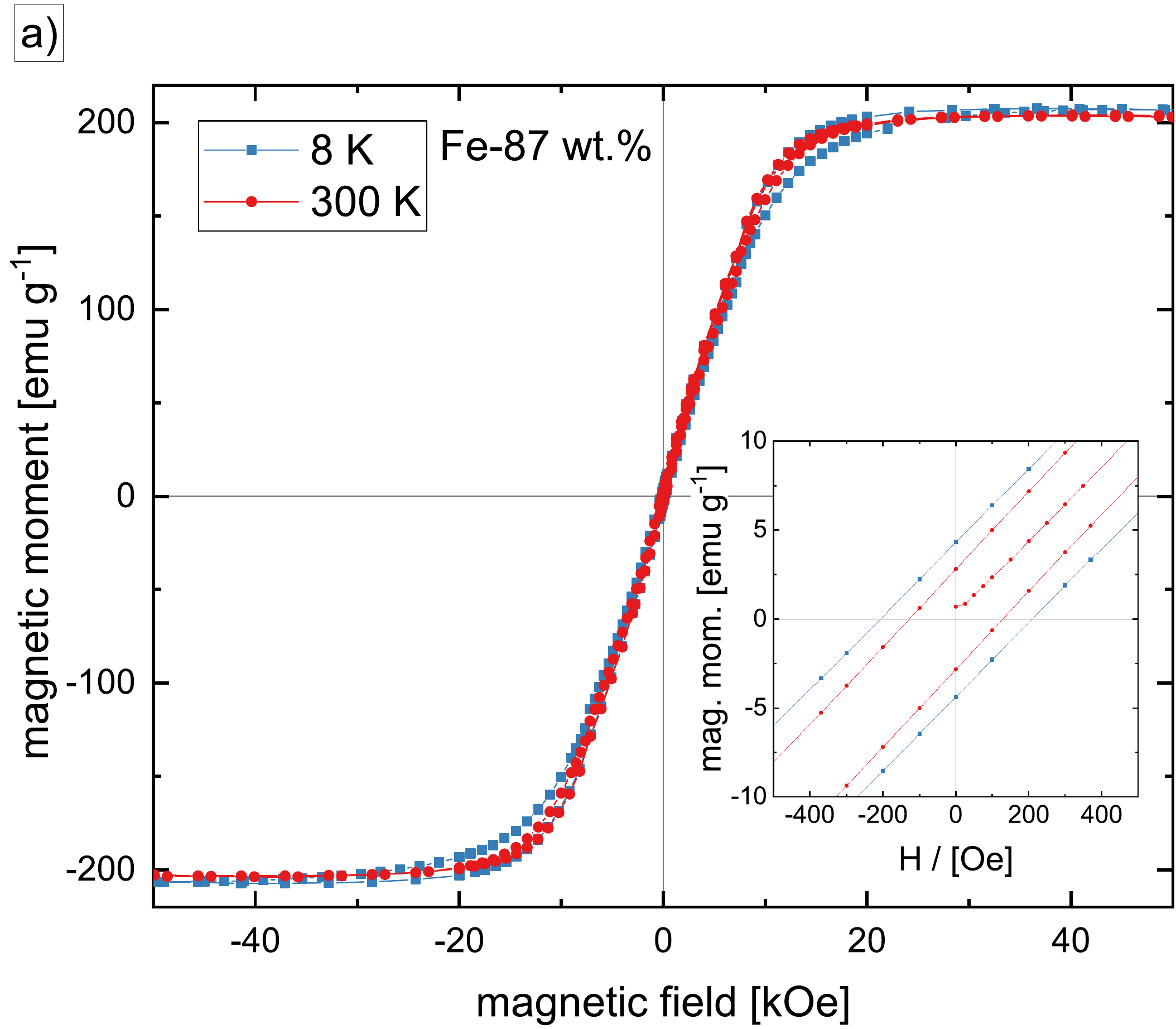}
	\includegraphics[width=0.4\textwidth]{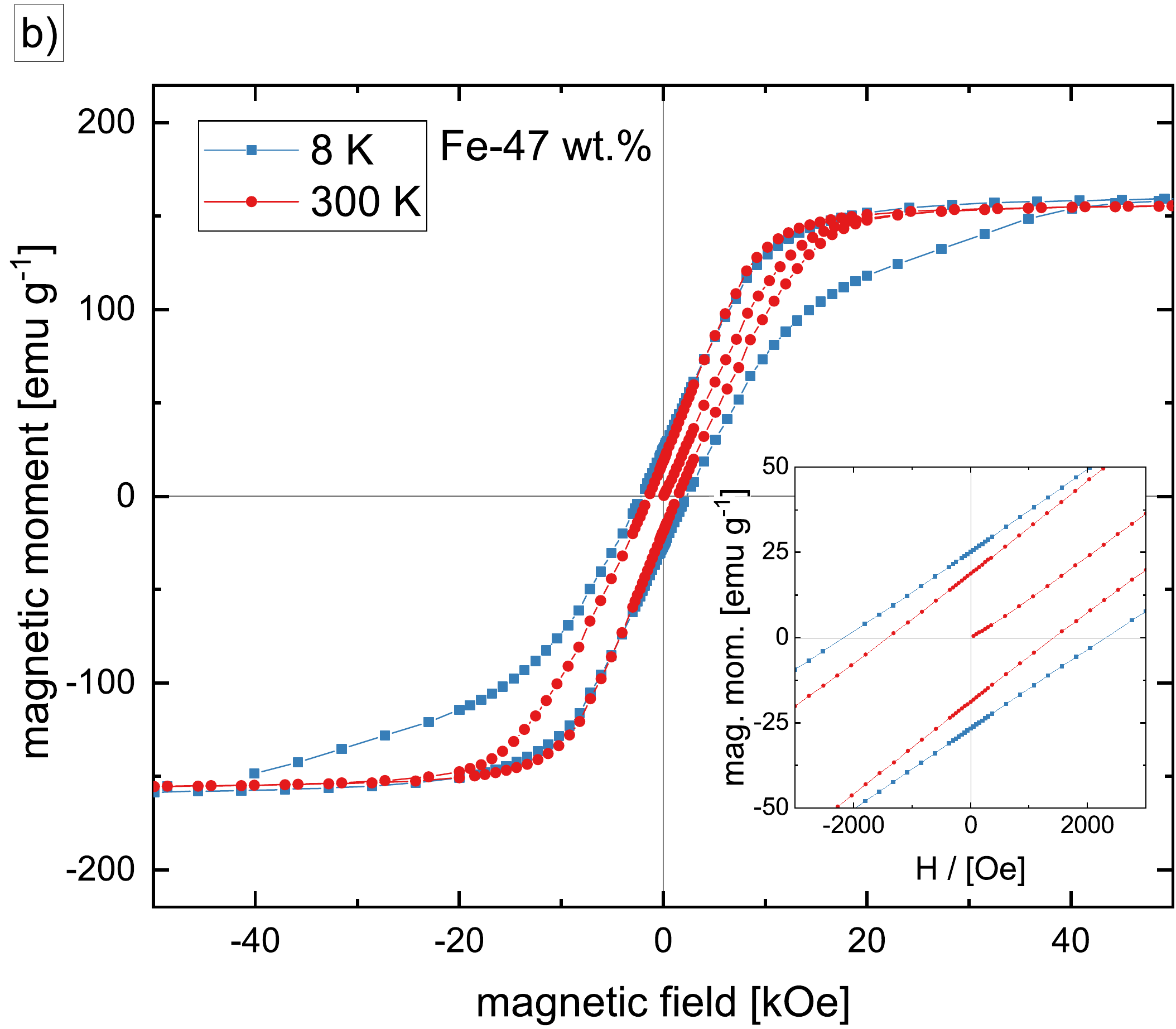}
	\includegraphics[width=0.4\textwidth]{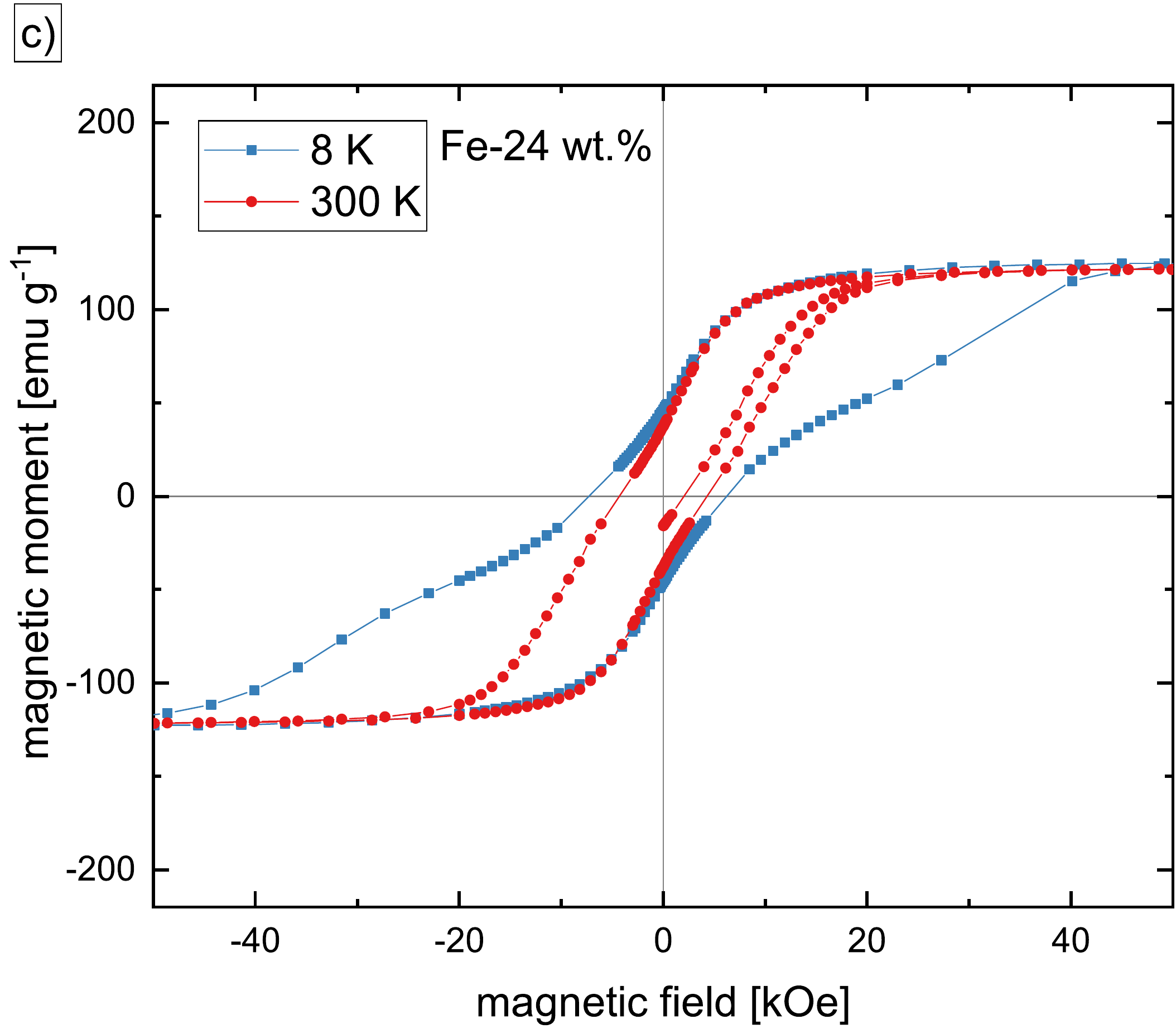}
	\includegraphics[width=0.4\textwidth]{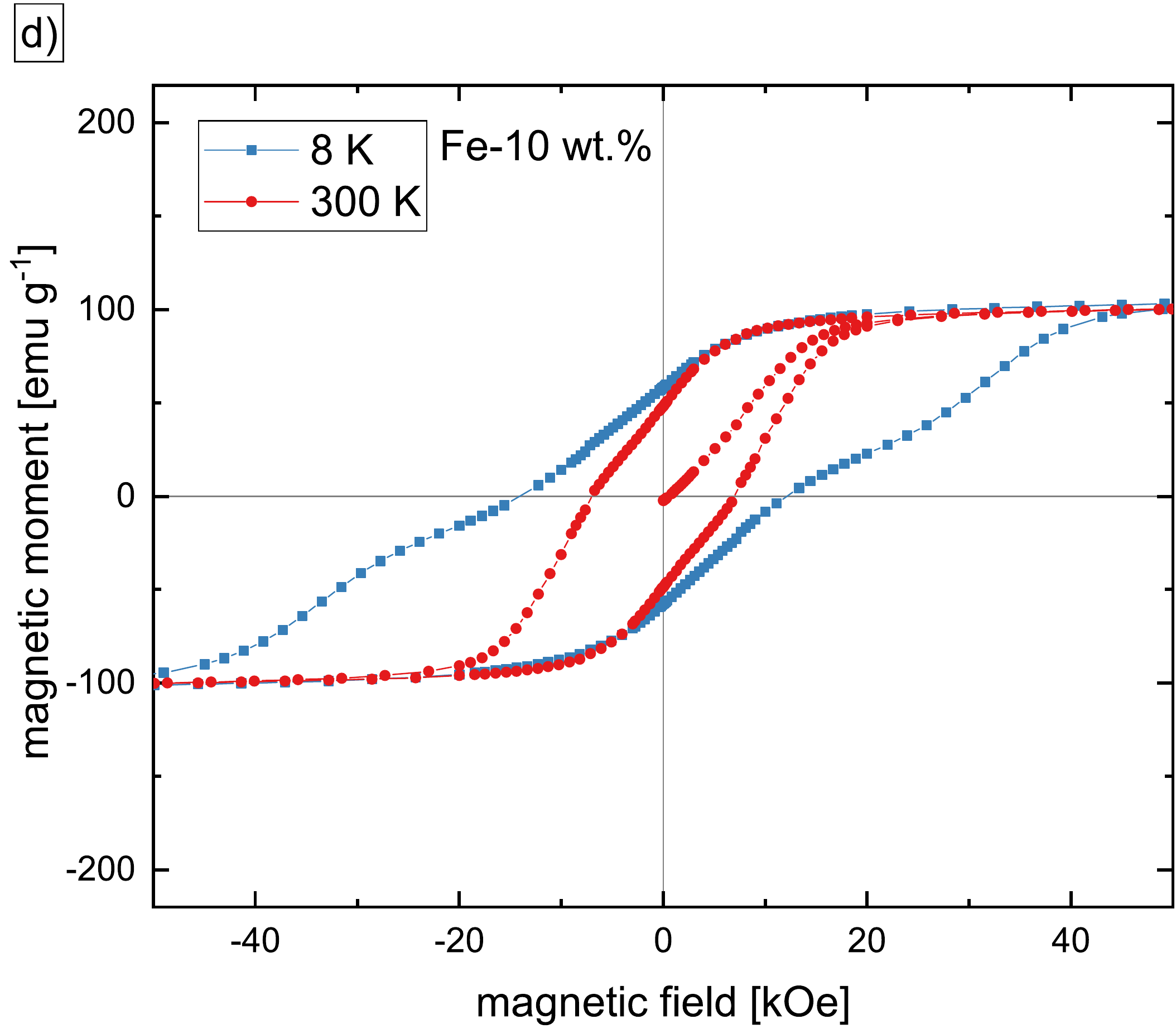}
	\caption{Hysteresis loops of HPT processed Fe-SmCo samples measured at 300~K (red, dots) and 8 K (blue, cubes) consisting of (\textbf{a}) 87, (\textbf{b}) 47, (\textbf{c}) 24 and (\textbf{d}) 10~wt.\% Fe. All diagrams are identically scaled, therefore insets in (\textbf{a},\textbf{b}) help to directly compare the coercivities at different temperatures.}
  \label{fig:SQUID,chemical}
\end{figure}

\subsubsection*{HPT-Induced Anisotropy}

The influence of the HPT induced microstructure on the magnetic behaviour is investigated by applying a magnetic field in different directions to the HPT disc for Fe-SmCo$_{5}$ samples with a Fe content of 24 and 47~wt.\% deformed at RT. The field is applied either in axial direction of the HPT-disc (as described above) or in tangential sample direction. The axial and tangential direction of the SQUID sample with respect to the HPT disc is visualized in Figure \ref{fig:schema} as well as in the inset of Figure \ref{fig:SQUID,anisotropy}. The respective directions are denoted as axial and tangential in the following.
The axial and tangential hysteresis of the selected samples consisting of 47 and 24~wt.\% Fe are shown in Figure \ref{fig:SQUID,anisotropy}a) and b). A higher susceptibility as well as a lower coercivity can be observed in the tangential direction, compared to the axial direction. A considerable shift of the tangential hysteresis to negative magnetic fields for the Fe-47~wt.\% sample and to positive fields for the Fe-24~wt.\% sample is visible. The difference between the positive and negative coercivities is $\sim$420~Oe and $\sim$710~Oe for the Fe-24~wt.\% and Fe-47~wt.\% sample, respectively.
While the axial hysteresis is measured first and finished at a maximum applied field of 70~kOe, the fully magnetised samples are mounted in tangential direction on the SQUID sample holder. Therefore, the direction of the shift is explained whether the sample is rotated up- or downwards.
Furthermore, the remanence is changed in both directions due to the shift.
Looking at Figure \ref{fig:SQUID,anisotropy}b), specific attention has to be given on the different saturation magnetisation values for the tangential hysteresis between positive and negative fields. As mentioned before, a completely magnetised hysteresis for the low Fe-content samples in axial direction cannot be assured. Due to the positive horizontal shift, the saturation magnetisation is not reached at positive fields whereas for decreasing fields, the shift assists a full magnetisation process. Hence, the measured values for the saturation magnetisation between applied negative and positive field of this hysteresis is different. 

\begin{figure}[H]
	\centering
	\includegraphics[width=0.4\textwidth]{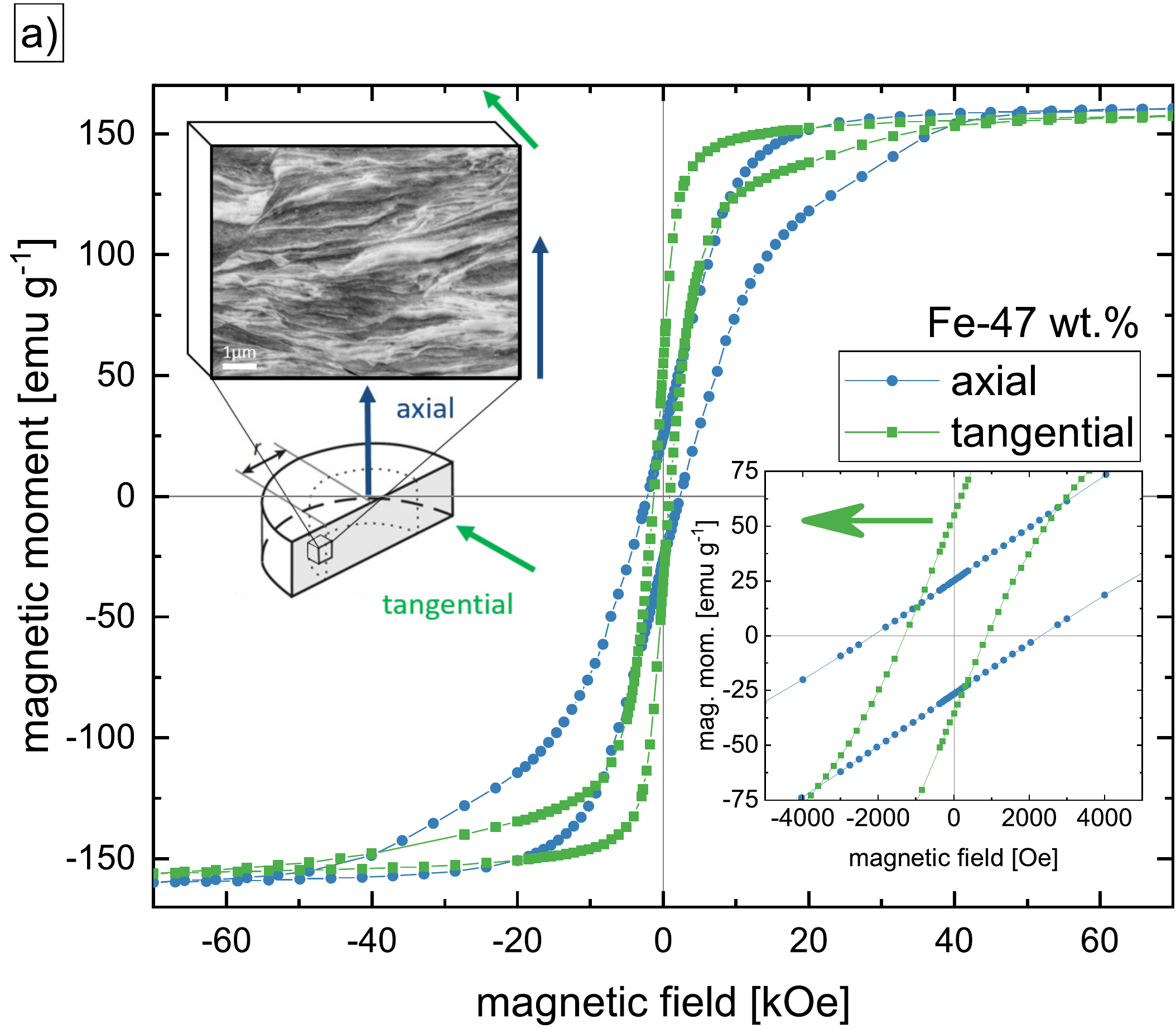}
	\includegraphics[width=0.4\textwidth]{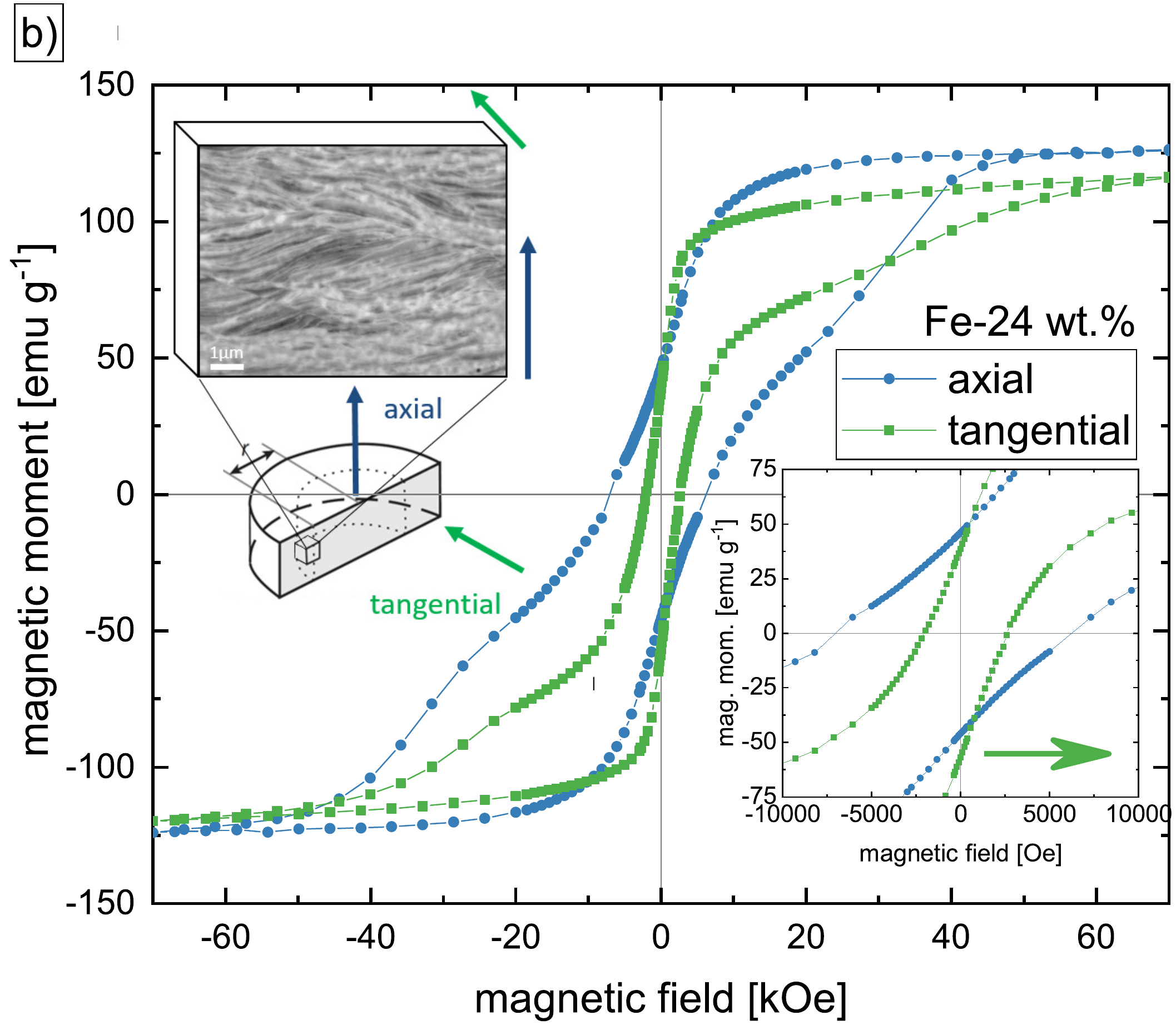}
	\caption{Hysteresis loops measured at 8 K in axial and tangential direction for Fe-SmCo$_{5}$ samples with (\textbf{a}) 24 and (\textbf{b}) 47 wt.\% Fe content. The inset shows the direction of the applied magnetic field compared to the HPT-disc.}
  \label{fig:SQUID,anisotropy}
\end{figure}

%%%%%%%%%%%%%%%%%%%%%%%%%%%%%%%%%%%%%%%%%%
\subsection{Strain-Dependent Magnetic Properties}

The influence of the applied strain on the magnetic properties is investigated for the Fe-SmCo$_{5}$ sample containing 47 wt.\% Fe--53 wt.\% SmCo$_{5}$. The strain is calculated according to Equation \ref{eq:gamma} for integer radii and the measured hysteresis correspond to different applied strains (Figure \ref{fig:SQUID,radial}). 
All samples have dimensions of several hundred $\mu$m in all spatial directions as shown in Figure \ref{fig:schema}. 
As a result, the sample preserves not a specific amount of applied strain, it rather varies as a linear relation of their radial position.
In the following the radius is used as sample indication and the reader is referred to Figure \ref{fig:microstructure_radial}, where the microstructure of the samples is shown. At r0, the spacing of the phases is larger than 50~$\upmu$m as can be seen in Figure \ref{fig:SQUID,radial}, which changes drastically when looking at r1. Here, the deformation process leads to a two-phase  microstructure where the SmCo$_{5}$ phase is fragmented and elongated, the spacing of the phases is significantly below 50~$\upmu$m in the axial direction. At r2 and r3, the morphology is further refined, where the axial dimension for the SmCo$_{5}$ phase is below 10~$\upmu$m at r2 and well below 1~$\upmu$m at r3. 
An exchange coupled hysteresis is found at 300 K  for all samples.
Therefore, it must be noticed again, that a relatively low applied strain for the sample at r0 corresponds to a radius between 0~<~r~$\leq$~0.5~mm (0~<~$\gamma$~<~20) and already results in an exchange coupled magnet. 
At r2 ($\gamma$~=~75) a maximum of the coercivity is found ($\sim$2000~Oe at 300~K and $\sim$3500~Oe at 8~K). 
Decoupled hysteresis are found at 8K, whereas the 'knee' is most pronounced for strains of $\gamma$ = 37 to 75. At a maximum strain of  $\gamma$~=~111, the appearance of the 'knee' as well as the coercivity decreases again.

\begin{figure}[H]
	\centering
	\includegraphics[width=0.4\textwidth]{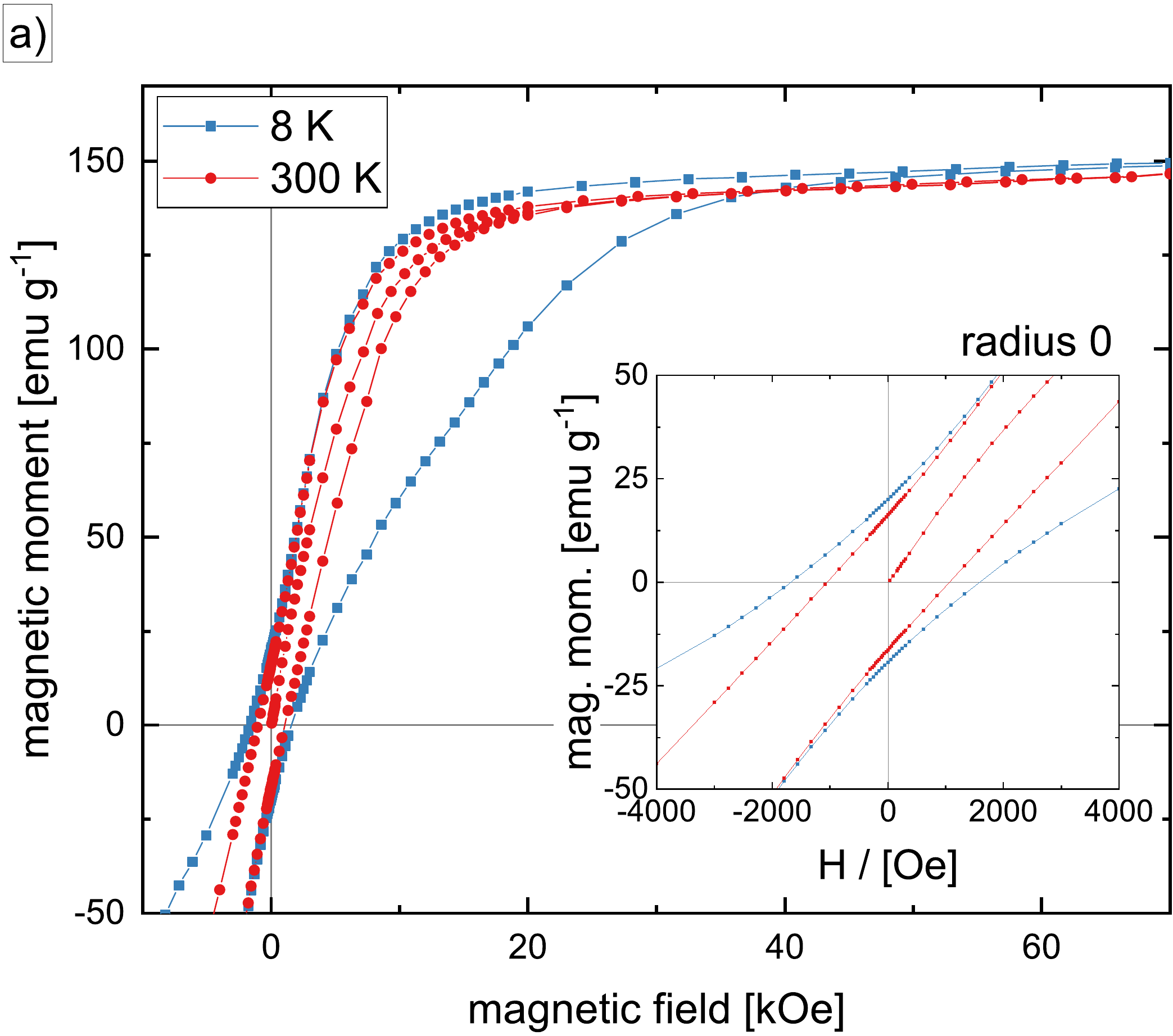}
	\includegraphics[width=0.4\textwidth]{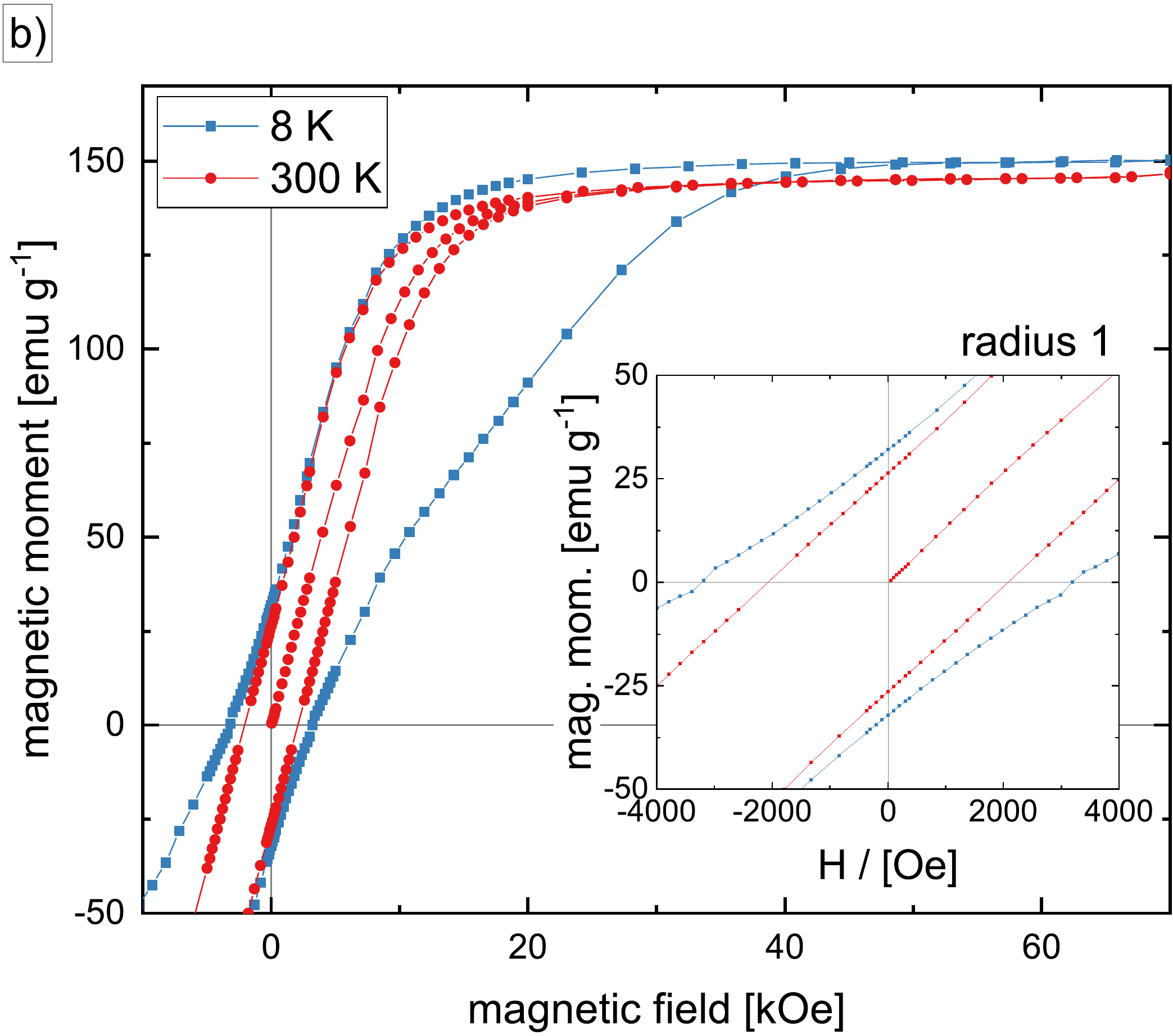}
	\includegraphics[width=0.4\textwidth]{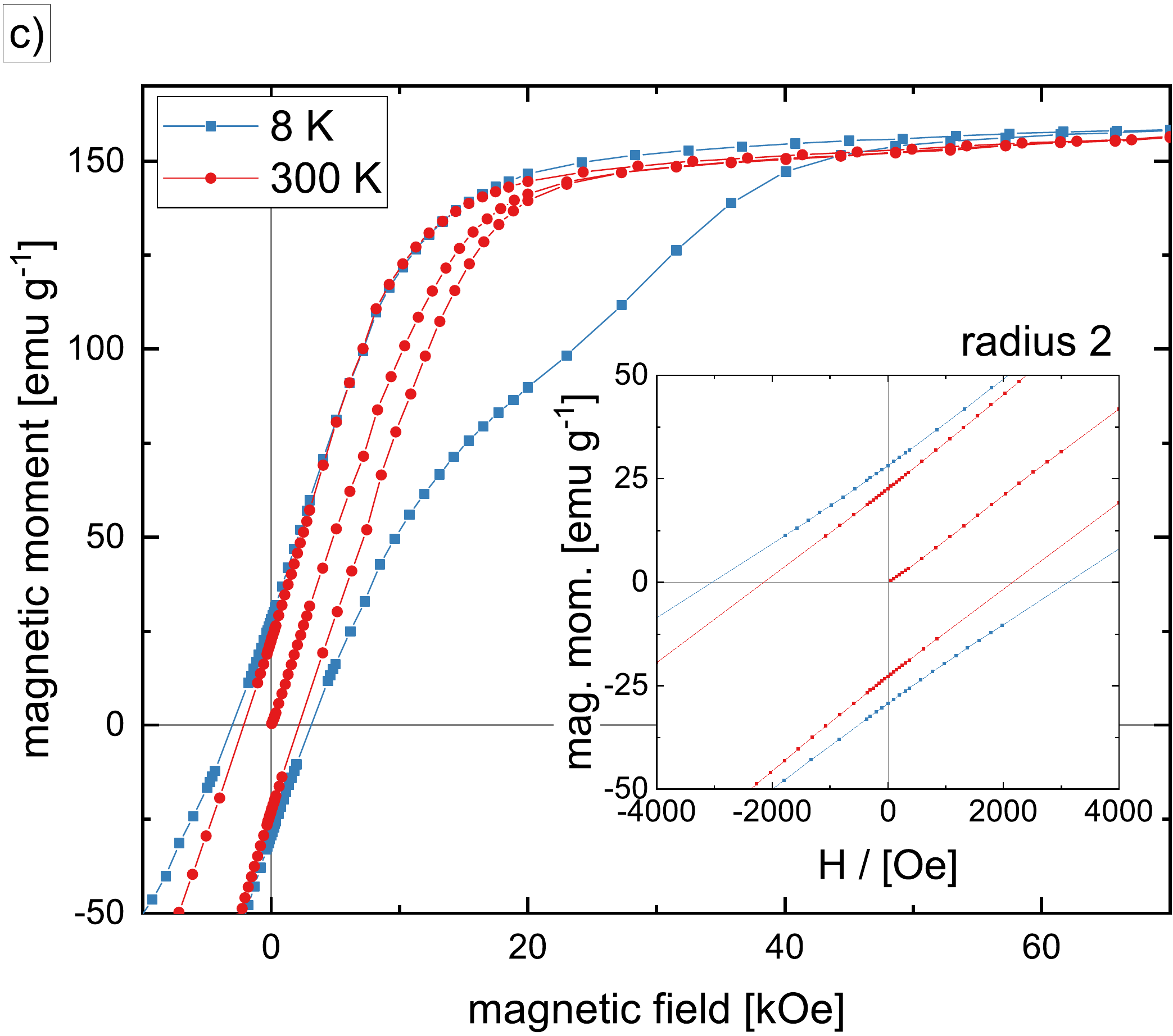}
	\includegraphics[width=0.4\textwidth]{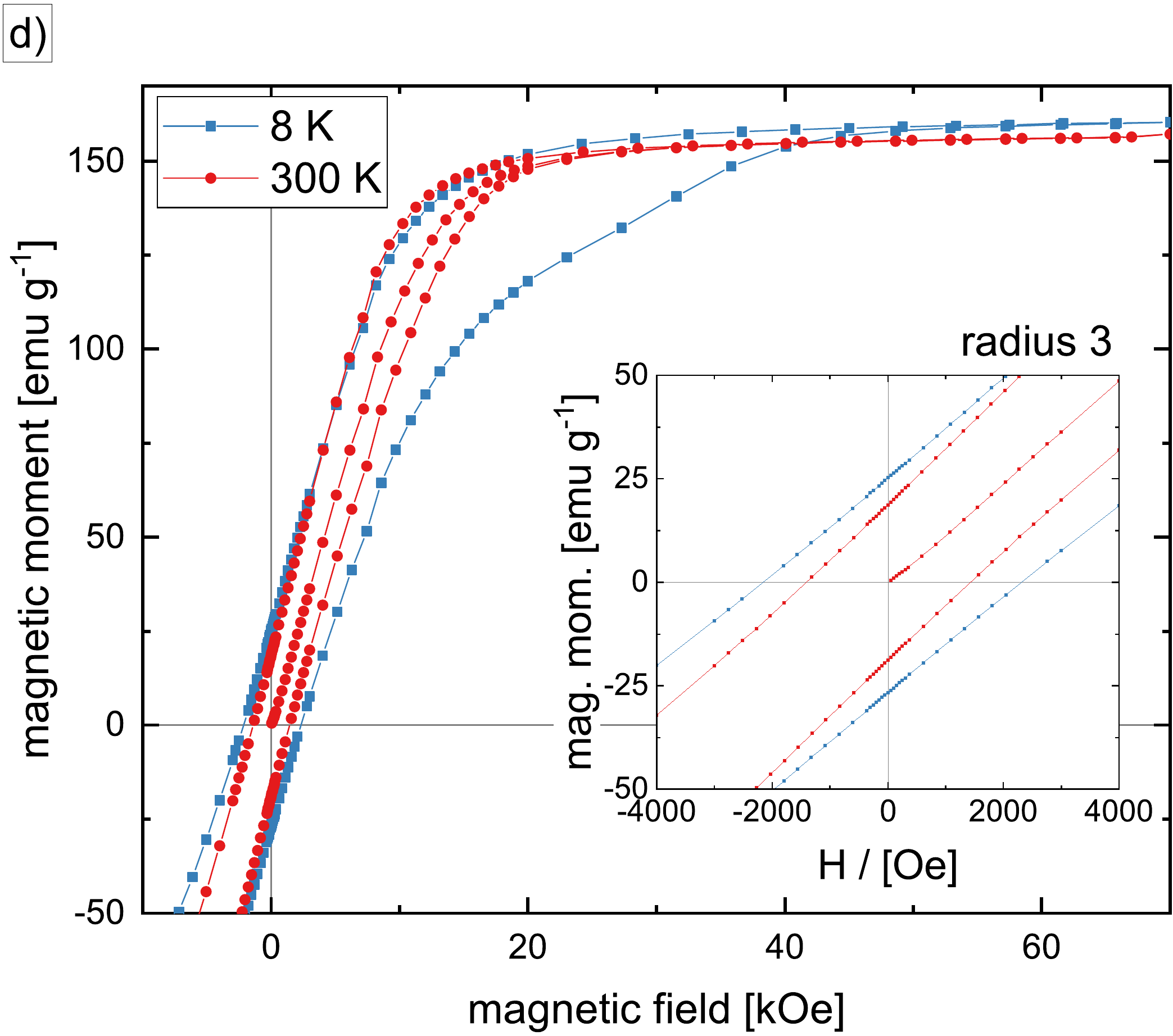}
	\caption{Hysteresis loops of HPT processed Fe-SmCo samples measured at 300 K (red, dots) and 8 K (blue, cubes) of a 47 wt.\% Fe--53 wt.\% SmCo$_{5}$ sample at a) r0, b) r1, c) r2 and d) r3.}
  \label{fig:SQUID,radial}
\end{figure}

%%%%%%%%%%%%%%%%%%%%%%%%%%%%%%%%%%%%%%%%%%
\section{Discussion}

In this work, Fe-SmCo$_{5}$ composites with different amounts of Fe are processed by HPT-deformation. SmCo$_{5}$ exhibits a hexagonal crystal structure, whereas $\alpha$-Fe is body centered cubic. During HPT-deformation, no ideal co-deformation of both phases takes place, Fe covers more strain than SmCo$_{5}$ particles. Nevertheless, both phases are refined resulting in an alternating finely dispersed phase morphology with a phase spacing in the 100~nm regime. 
Compared to a cubic crystal, hexagonal structures provide less sliding planes and are more prone to localized shear and brittle behaviour. 
The deformation of hexagonal crystal structures is usually limited by pronounced crack formation or accompanied by phase transformations \cite{kormout2017, bachmaier2017, edalati2016, wang2019, stueckler2019}. Especially at low strains, the SmCo$_{5}$ phase shows a lack of phase deformation and the formation of cracks in this phase is visible. 
Nonetheless, fragmentation and phase refinement is observed. However, most of the plastic deformation is localized in the Fe phase, where grain refinement as in single phase HPT deformed Fe is expected.
When the composites are exposed to a higher amount of strain, the SmCo$_{5}$ phase is plastically deformed to a higher extend.
For the Fe phase, a grain refinement with increasing strain is observed instead.  
It is assumed, that fracturing and localized shear of the brittle and mechanically harder SmCo$_{5}$ phase dominates the deformation behaviour until the hardness of the Fe and SmCo$_{5}$ phase converges due to the grain refinement of the Fe phase. Then, the SmCo$_{5}$ phase provides a higher plasticity \cite{kiparisov1987}. However, local mutual hardness differences might lead to a partial co-deformation or an alternating deformation of the two phases. The rheologically harder phase tends to local necking and breaks into smaller fragments leading to the even more refined structure at the highest strains \cite{chen2001}.
However, the possibility to deform hexagonal SmCo$_{5}$ is noticeable.

For exchange coupled magnets two magnetic phases are necessary. Thereby the hard magnetic material is the main phase, providing a high anisotropy and coercivity. The addition of a soft magnetic phase reduces the anisotropy and coercivity, but enhances the magnetisation and remanence \cite{skomski2013}. Depending on morphology (e.g., spheres, layers, cylinders) of the dispersion and used hard and soft magnetic materials, a maximised theoretical energy product can be achieved. The energy product usually peaks around 30~wt.\% soft magnetic material, which emphasizes the importance of a hard magnetic material matrix, which encloses the soft magnetic phase. Another crucial parameter is the exchange length for the soft magnetic material. Because of the small magnetic anisotropy constant of the soft phase, the spin orientation of the hard phase induces an exchange coupling into the soft phase. At the interface the spins of the soft phase align with the hard phase. This usually holds if the dimensions of the soft phase do not exceed a range of twice the exchange length 

\begin{equation}\label{eq:exchange-length}
l_{ex}^{soft} = \sqrt{\dfrac{A^{soft}}{K^{soft}_{1}}}
\end{equation}
where \textit{A} denotes the exchange constant and \textit{K} the anisotropy constant (for $\alpha$-Fe: A = 2.5  $\times$  10$^{-11}$ Jm$^{-1}$, K$_{1}$ = 4.6 $\times$ 10$^{4}$ Jm$^{-3}$, l$_{ex}$ = 23 nm) \cite{kronmueller1996, kneller1991}. 
SEM investigations (Figures \ref{fig:microstructure_radial}, \ref{fig:microstructure,chemical} and \ref{fig:microstructure,linescan}a) show a fine phase morphology in the 100~nm regime, sustained by a peak broadening in XRD experiments (Figure \ref{fig:microstructure,linescan}b). The dimensions of the phases being small enough, are confirmed by the measured magnetic properties. As depicted in Figure~\ref{fig:SQUID,chemical}, completely exchange coupled hysteresis with a concave shape are found. The results are comparable to magnetic measurements received of ball milled Fe-SmCo powders \cite{rong2011}.

By analysing the virgin curves presented in Figure \ref{fig:SQUID,chemical}, a material classification between nucleation controlled and pinning controlled magnets can be given \cite{strnat1991}. While existing domains are easily expanded in nucleation controlled magnets, microstructural obstacles (e.g., dislocations or vacancies, such as they occur after HPT deformation) impede the movement of domain walls, which act as pinning centers. Therefore, the initial susceptibility is lower and the saturation magnetisation is reached at higher fields indicating a pinning controlled magnetic material after HPT-deformation.

The hysteresis shapes at 8~K  strongly differ from the 300 K curves, where exchange coupling is found. Hysteresis curves showing such a 'knee' at low temperatures are suggested as a decoupled compound, when hard and soft magnetic phases do not interact and are measured as magnetically independent phases~\cite{kneller1991}. 
The effect, that decoupling at decreasing temperatures occurs, can be explained by a temperature dependence of the magnetocrystalline anisotropy of the hard SmCo$_{5}$ phase \cite{sankar1975}. As the anisotropy increases with decreasing temperature, according to Equation (\ref{eq:exchange-length}) the exchange length decreases \cite{liu2000}. Therefore, the ratio between total volume and exchange coupled volume diminishes at lower temperatures \cite{munoz2012}.

The shapes of tangential and axial hysteresis in  Figure \ref{fig:SQUID,anisotropy}b) are completely different. In axial direction a sheared and decoupled hysteresis is visible. The tangential curve offers a higher susceptibility with nearly no decoupling characteristics. 
This behaviour is similar for the hysteresis in Figure~\ref{fig:SQUID,anisotropy}a) besides a stronger pronounced decoupling due to the higher SmCo$_{5}$ content.
Both tangential hysteresis are horizontally shifted which can be explained by the microstructure and the magnetic nucleation field. 
In the literature, a texture formation of hexagonal crystal structures is often observed during HPT-deformation. Therefore, the c-axis orientates into the shearing or tangential direction \cite{wang2019, bonarski2008, huang2013}. Furthermore, the c-axis also represents the magnetically easy axis, which means that all SmCo$_{5}$ crystals are easily demagnetised. In the case of the used SmCo$_{5}$ material, this also explains the higher susceptibility compared to the axial hysteresis. 
Hence, we assume that the microstructure of HPT-deformed Fe-SmCo$_{5}$ composites consists of a lamellar-like structure with the easy axis of the hard magnetic phase oriented in tangential direction. The soft magnetic spins align in the same direction. When the field is reversed areas with opposite spins have to be formed. This requires a so called nucleation field H$_{n}$ and is described by

\begin{equation}
H_n \cong \dfrac{K^{hard}}{\mu_0 M_s^{soft}}
\end{equation}
where \textit{K} is the anisotropy constant and \textit{M$_{s}$} the saturation magnetisation \cite{kneller1991, guo2002}. If the hysteresis cannot be demagnetised completely, a pinning between hard and soft magnetic spins remains. The nucleation field is changed due to a varying magnetisation and the material is demagnetised at different fields.

In the literature, ball milling experiments, as a processing technique using plastic deformation, are mainly used to fabricate Fe-SmCo$_{5}$ magnets. Although the production technique differs a lot, comparisons of the results can be made. 
Rong et al. \cite{rong2011} reported a decrease in coercivity when milling the powder up to 10~h. They describe the formation of an amorphous structure, due to a reduced hard-phase stability. Further, they found an increasing saturation magnetisation because of the formation of a FeCo-alloy providing a higher magnetisation compared to $\alpha$-Fe. 
During HPT also a strong decrease in coercivity and a slight increase in the magnetisation is found when the applied strain of the HPT-deformed sample exceeds $\gamma$~=~75. 
Besides a slight peak broadening, no amorphization is found in our corresponding synchrotron XRD measurements (Figure \ref{fig:microstructure,linescan}b), which could explain the coercivity deterioration. 
Detailed TEM studies at the interfaces between the Fe-SmCo$_{5}$ interfaces could reveal the formation of a small non-magnetic layer in between, if a slight decomposition of the SmCo$_{5}$ structure is present.
However, this is beyond the scope of this work, which aims to show that HPT-deformation is an alternative process to manufacture exchange coupled nanocomposites. 

Therefore, future investigations will cover the influence of applied amount of strain on the magnetic behaviour for different chemical compositions and other phases. For example, the magnetic properties will be enhanced when Fe is substituted by FeCo alloys and SmCo$_{5}$ by other SmCo based intermetallic phases such as Sm$_{2}$Co$_{7}$ or Sm$_{2}$Co$_{17}$.
The HPT-parameters, such as a different deformation temperature, will be adjusted and could lead to a better deformation performance, while the phases remain stable. In the present study it could be shown that Fe and SmCo$_{5}$ phases remain stable until 250 $^{\circ}$C
 (see Figure \ref{fig:DESY,temperature}a). 
Moreover, the stability of the finely dispersed heterostructure of the magnets will be tested by annealing at higher temperatures to show the utilization for technical applications. Finally, subsequent annealing treatments at low temperatures after deformation are expected to reduce crystal defects, introduced by the shearing process, which again is expected to improve magnetic performance.

%%%%%%%%%%%%%%%%%%%%%%%%%%%%%%%%%%%%%%%%%%
\section{Conclusions}

For the first time, the possibility of producing nanostructured Fe-SmCo exchange spring magnets by HPT-deformation over a broad chemical composition range is demonstrated. HPT-deformation at elevated temperatures (T~>~250 $^{\circ}$C
) leads to a strain induced phase transformation of the SmCo-phase, whereas room temperature deformation allows the production of a defined heterogeneous nanostructured morphology preserving the soft-magnetic Fe phase and hard magnetic SmCo$_{5}$-phase. In contrast to other techniques, it is possible to fabricate nanostructured bulk samples already exhibiting dimensions of several centimetres without any thermal compaction or sintering process. 
Spring magnetic behaviour at 300~K and a decoupling effect at lower temperatures (8~K) are found for all chemical compositions. 
HPT-deformation leads to an anisotropic microstructure, which is magnetically investigated for two different chemical compositions in axial and tangential HPT-disc directions. 
The influence of the applied strain on the magnetic properties is investigated on the Fe-SmCo$_{5}$ sample with medium composition, revealing an optimum amount of strain of $\gamma$~=~75, for the used initial powder size, showing the highest coercivity. 

%%%%%%%%%%%%%%%%%%%%%%%%%%%%%%%%%%%%%%%%%%%
%\section{Patents}
%This section is not mandatory, but may be added if there are patents resulting from the work reported in this manuscript.

%%%%%%%%%%%%%%%%%%%%%%%%%%%%%%%%%%%%%%%%%%
\vspace{6pt} 

%%%%%%%%%%%%%%%%%%%%%%%%%%%%%%%%%%%%%%%%%%
%% optional
%\supplementary{The following are available online at \linksupplementary{s1}, Figure S1: title, Table S1: title, Video S1: title.}

% Only for the journal Methods and Protocols:
% If you wish to submit a video article, please do so with any other supplementary material.
% \supplementary{The following are available at \linksupplementary{s1}, Figure S1: title, Table S1: title, Video S1: title. A supporting video article is available at doi: link.}

%%%%%%%%%%%%%%%%%%%%%%%%%%%%%%%%%%%%%%%%%%
\textbf{Author Contributions:} Conceptualization, L.W. and A.B.; Methodology, L.W., S.W. and A.B.; Validation, L.W., M.S., S.W., P.K., H.K., R.P. and A.B.; Formal Analysis, L.W.; Investigation, L.W.; Data Curation, L.W.; Writing---Original Draft Preparation, L.W.; Writing---Review and Editing, R.P., H.K., P.K., M.S., S.W., and A.B.; Visualization, L.W.; Supervision, H.K., P.K., R.P. and A.B.; Project Administration, A.B.; Funding Acquisition, A.B. All authors have read and agreed to the published version of the manuscript.

%%%%%%%%%%%%%%%%%%%%%%%%%%%%%%%%%%%%%%%%%%
\textbf{Funding:} This project has received funding from the European Research Council (ERC) under the EuropeanUnion’s Horizon 2020 research and innovation programme (Grant No. 757333).

%%%%%%%%%%%%%%%%%%%%%%%%%%%%%%%%%%%%%%%%%%
\textbf{Acknowledgments:} The measurements leading to these results have been performed at PETRA III: P07 at DESYHamburg (Germany), a member of the Helmholtz Association (HGF). We gratefully acknowledge the assistance by Norbert Schell and further thank F. Spieckermann and C. Gammer for their help with data processing. L.W. deeply appreciates the help of Mirjam Spuller and Georg Holub for specimen preparation.

%%%%%%%%%%%%%%%%%%%%%%%%%%%%%%%%%%%%%%%%%%
\textbf{Conflicts of Interest:} The authors declare no conflict of interest.

%%%%%%%%%%%%%%%%%%%%%%%%%%%%%%%%%%%%%%%%%%
%% optional
\textbf{Abbreviations}\\
The following abbreviations are used in this manuscript:\\

\noindent 
\begin{tabular}{@{}ll}
BSE		&	Backscattered Electrons \\
COD		&	Crystallography Open Database\\
EDX		&	Energy Dispersive X-ray Spectroscopy\\
HPT		&	High-Pressure Torsion\\
RT		&	Room Temperature\\
SEM		&	Scanning Electron Microscopy\\
SPD		&	Severe Plastic Deformation\\
SQUID	&	Superconducting Quantum Interference Device\\
TEM		&	Transmission Electron Microscopy
	
\end{tabular}

\appendix
\section{Radial Microstructure} \label{sec:appendix}

\begin{figure}[H]
\centering
\includegraphics[width=10 cm]{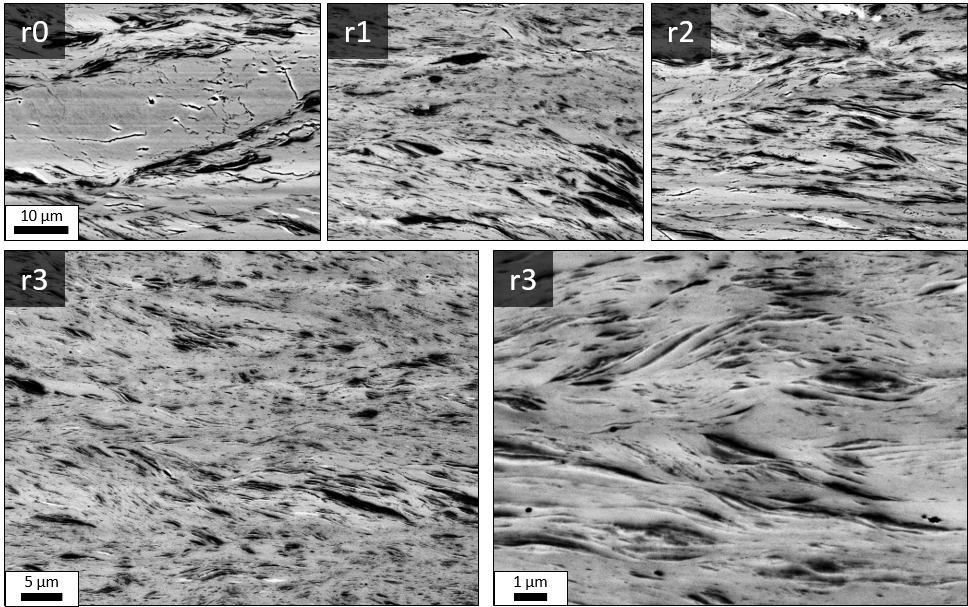}
\caption{BSE images of a HPT-deformed sample (Nr. 1) at RT consisting of 10 wt.\% Fe--90 wt.\% SmCo$_{5}$ at different radii. The scale bar in r0 also applies to r1 and r2. Images of two magnifications are presented for r3.}
  \label{fig:microstructure_Appendix_01}
\end{figure}

\begin{figure}[H]
\centering
\includegraphics[width=13 cm]{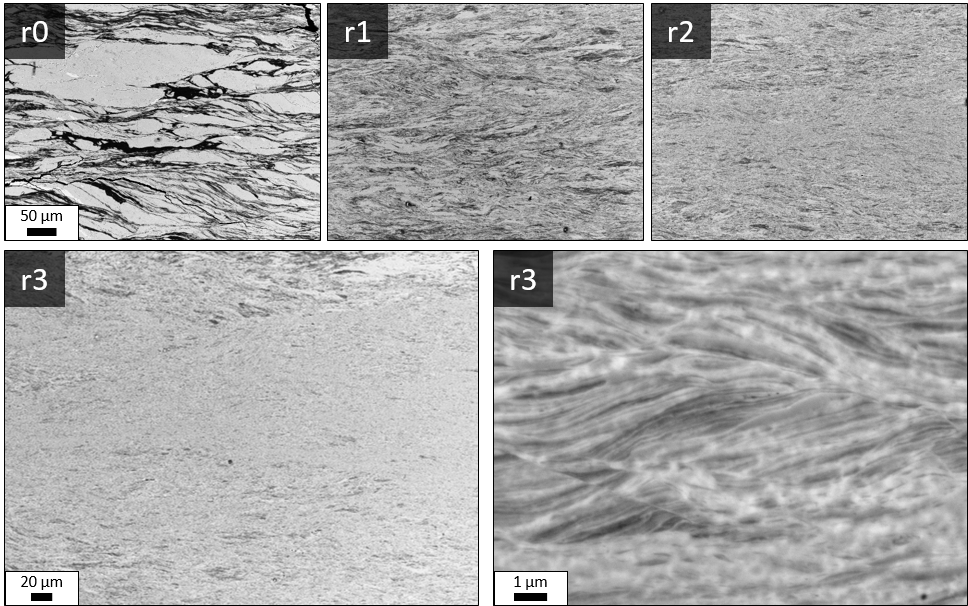}
\caption{BSE images of a HPT-deformed sample (Nr. 2) at RT consisting of 24 wt.\% Fe--76 wt.\% SmCo$_{5}$ at different radii. The scale bar in r0 also applies to r1 and r2. Images of two magnifications are presented for r3.}
  \label{fig:microstructure_Appendix_02}
\end{figure}

\begin{figure}[H]
\centering
\includegraphics[width=13 cm]{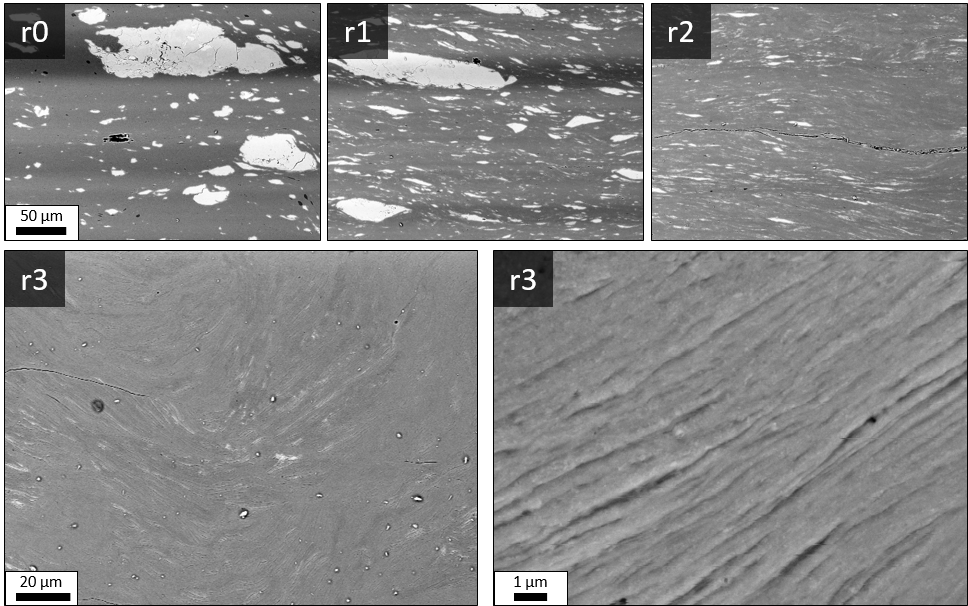}
\caption{BSE images of a HPT-deformed sample (Nr. 4) at RT consisting of 87 wt.\% Fe--13 wt.\% SmCo$_{5}$ at different radii. The scale bar in r0 also applies to r1 and r2. Images of two magnifications are presented for r3.}
  \label{fig:microstructure_Appendix_04}
\end{figure}

\begin{figure}[H]
\centering
\includegraphics[width=14 cm]{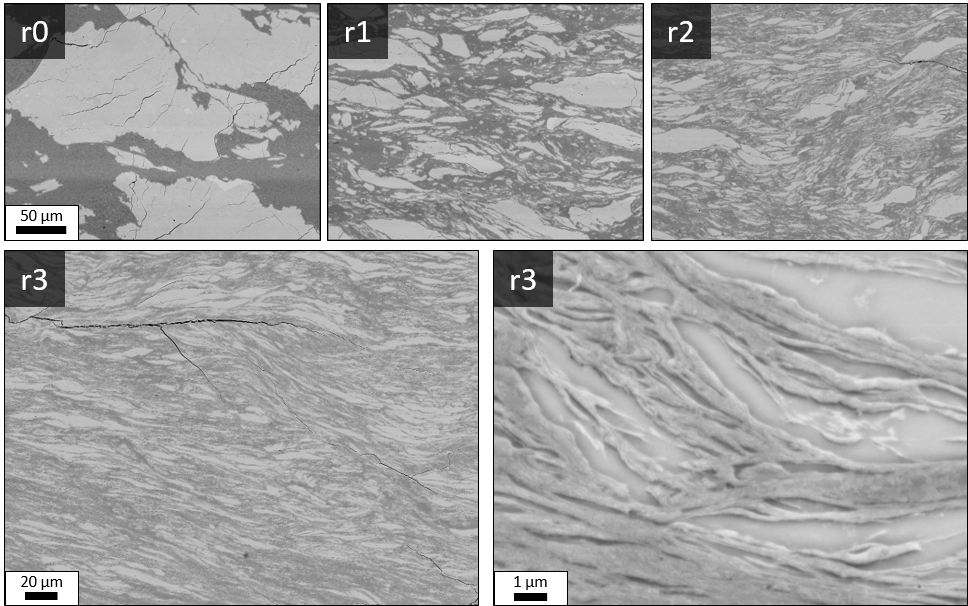}
\caption{BSE images of a HPT-deformed sample (Nr. 5) at 250 $^{\circ}$C
 consisting of 47 wt.\% Fe--53 wt.\% SmCo$_{5}$ at different radii. The scale bar in r0 also applies to r1 and r2. Images of two magnifications are presented for r3.}
  \label{fig:microstructure_Appendix_05}
\end{figure}

\begin{figure}[H]
\centering
\includegraphics[width=14 cm]{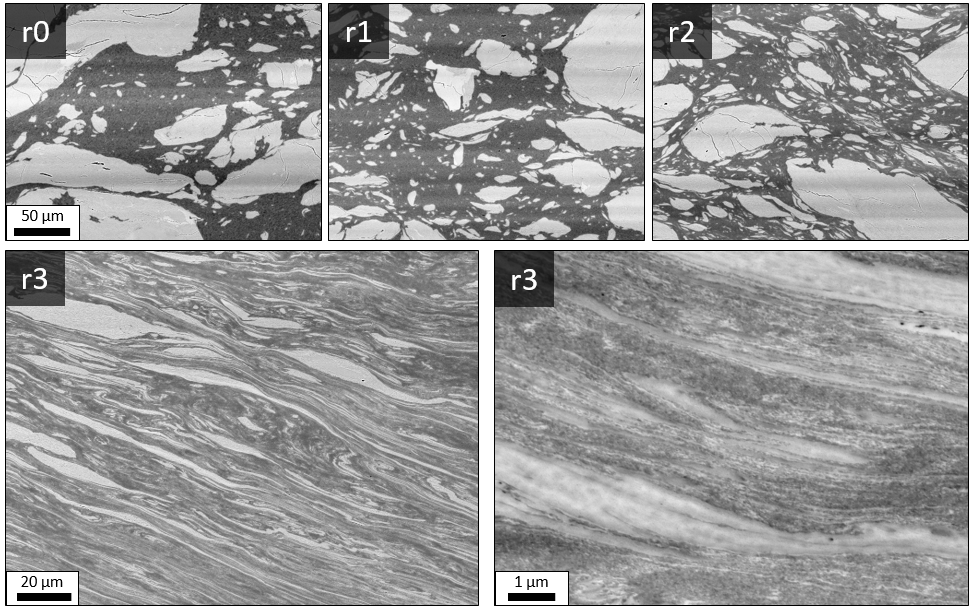}
\caption{BSE images of a HPT-deformed sample (Nr. 6) at 400 $^{\circ}$C
 consisting of 47 wt.\% Fe--53 wt.\% SmCo$_{5}$ at different radii. The scale bar in r0 also applies to r1 and r2. Images of two magnifications are presented for r3.}
  \label{fig:microstructure_Appendix_06}
\end{figure}

In this section, supplementary information on the processed samples is given. 
The change of the microstructure as a function of applied strain is described in detail for sample Nr. 3 in Section \ref{sec:microstructure}. Similarities are found for sample Nr. 1, 2 and 4, depicted in Figures \ref{fig:microstructure_Appendix_01}, \ref{fig:microstructure_Appendix_02} and \ref{fig:microstructure_Appendix_04}, respectively.
The radial microstructure for samples deformed at elevated temperatures is found in Figures \ref{fig:microstructure_Appendix_05} (250 $^{\circ}$C
) and \ref{fig:microstructure_Appendix_06} (400 $^{\circ}$C
).

%%%%%%%%%%%%%%%%%%%%%%%%%%%%%%%%%%%%%%%%%%
\bibliography{literature_FeSmCo}

\end{document}